\documentclass[journal,final]{IEEEtran}
\usepackage[T1]{fontenc}
\usepackage[latin9]{inputenc}
\usepackage{amsmath}
\usepackage{graphicx}
\usepackage{amssymb}

\makeatletter

\makeatletter

\usepackage{amsfonts}

\usepackage{epsfig}

\usepackage{mathrsfs}

\usepackage[scriptsize]{subfigure}

\newcounter{rem}
\newcounter{algo}
\newenvironment{algo}[2]{\refstepcounter{algo}\label{#2}   \begin{center}
\begin{minipage}{0.48\textwidth}   \hrule\smallskip
\textbf{Algorithm \thealgo: #1}
\par\smallskip\hrule\smallskip\ignorespaces}{\par\smallskip\hrule
\end{minipage}
\end{center}
}

\ifCLASSINFOpdf
\else
\fi
\newtheorem{theorem}{Theorem}
\newtheorem{proposition}{Proposition}
\newtheorem{definition}{Definition}
\newtheorem{lemma}{Lemma}
\newtheorem{corollary}{Corollary}

\newenvironment{proof}[1][Proof]{\noindent \textbf{#1.} }{\qedsymbol}
\newcommand{\qedsymbol}{\hspace{\fill}\rule{1.5ex}{1.5ex}}
\input{tcilatex}

\makeatother

\begin{document}

\title{Competitive Design of Multiuser MIMO Systems based on Game Theory:
A Unified View }

\author{Gesualdo~Scutari,~\IEEEmembership{Member,~IEEE,} Daniel~P.~Palomar,~\IEEEmembership{Member,~IEEE,}
and~Sergio~Barbarossa,~\IEEEmembership{Member,~IEEE}
\thanks{G. Scutari and S. Barbarossa are with the INFOCOM Department, University
of Rome, {}``La Sapienza,\textquotedblright Rome, Italy. E-mails:
\texttt{$\{$scutari, sergio$\}$@infocom.uniroma1.it}.%
}
\thanks{D. P. Palomar is with the Department of Electronic and Computer Engineering,
Hong Kong University of Science and Technology, Hong Kong. E-mail:
\texttt{palomar@ust.hk}.%
}
\thanks{Manuscript received August 15, 2007; revised May 10, 2008.%
}%
\thanks{{}%
}}

\maketitle
\markboth{Journal of Selected Areas in Communications,~Vol.~?, No.~?, September~2008}{Shell
\MakeLowercase{\textit{et al.}}: Bare Demo of IEEEtran.cls for
Journals} 


\begin{abstract}
This paper considers the noncooperative maximization of mutual information
in the Gaussian interference channel in a fully distributed fashion
via game theory. This problem has been studied in a number of papers
during the past decade for the case of frequency-selective channels.
A variety of conditions guaranteeing the uniqueness of the Nash Equilibrium
(NE) and convergence of many different distributed algorithms have
been derived. In this paper we provide a unified view of the state-of-the-art
results, showing that most of the techniques proposed in the literature
to study the game, even though apparently different, can be unified
using our recent interpretation of the waterfilling operator as a
projection onto a proper polyhedral set. Based on this interpretation,
we then provide a mathematical framework, useful to derive a unified
set of sufficient conditions guaranteeing the uniqueness of the NE
and the global convergence of waterfilling based asynchronous distributed
algorithms.

The proposed mathematical framework is also instrumental to study
the extension of the game to the more general MIMO case, for which
only few results are available in the current literature. The resulting
algorithm is, similarly to the frequency-selective case, an iterative
asynchronous MIMO waterfilling algorithm. The proof of convergence
hinges again on the interpretation of the MIMO waterfilling as a matrix
projection, which is the natural generalization of our results obtained
for the waterfilling mapping in the frequency-selective case. 
\end{abstract}


\IEEEpeerreviewmaketitle

\section{Introduction}

\IEEEPARstart{T}{he} interference channel is a mathematical model
relevant to many physical communication channels and multiuser systems
where multiple uncoordinated links share a common communication medium,
such as digital subscriber lines \cite{Starr-Cioffi Book}, single
(or multi) antenna cellular radio, ad-hoc wireless networks \cite{Goldsmith-Wicker,Akyildiz-Wang},
and cognitive radio systems \cite{Haykin}.

The interference channel is characterized by its capacity region,
defined as the set of rates that can be simultaneously achieved by
the users in the system while making the error probability arbitrary
small. A pragmatic approach that leads to an achievable region or
inner bound of the capacity region is to restrict the system to operate
as a set of independent units, i.e., not allowing multiuser encoding/decoding
or the use of interference cancellation techniques. This approach
is very relevant in practical systems, as it limits the amount of
signaling among the users. With this assumption, multiuser interference
is treated as additive colored noise and the system design reduces
to finding the optimum covariance matrix of the symbols transmitted
by each user.

Within this context, in this paper we consider the maximization of
mutual information in a fully distributed fashion using a game theoretical
approach. Since the seminal paper of Yu et al. \cite{Yu} in 2002
(and the conference version in 2001), this problem has been studied
in a number of works during the past seven years for the case of \emph{SISO}
\emph{frequency-selective channels} or, equivalently, a set of parallel
non-interfering scalar channels \cite{ChungISIT03}-\cite{Scutari-IT-08}.
In the cited papers, the maximization of mutual information is formulated
as a strategic noncooperative game, where every SISO link is a player
that competes against the others by choosing his power allocation
(transmission strategy) over the frequency bins (or parallel channels)
to maximize his own information rate (payoff function).%
\footnote{The choice of the gender of the players is always controversial in
the literature of game theory. This is reflected mainly on the use
the third-person singular pronouns: some authors use {\small {}``}his\char`\"{},
while others use {\small {}``}her{\small \textquotedblright}, and
others--more diplomatic ones--even use {\small {}``}his/her{\small \textquotedblright}.
English non-native speakers tend to use {\small {}``}its{\small \textquotedblright}
to avoid the problem but that is not well accepted by native speakers.
In two-player zero-sum games, the issue is even trickier and some
authors resort to the use of one gender for the good player and another
for the bad player. See the foreword in \cite{Osborne} for a related
discussion on the issue. In this paper, for simplicity of notation
and without further implications, we simply use {\small {}``}his{\small \textquotedblright}.%
} Based on the celebrated notion of Nash Equilibrium (NE) in game theory
(cf. \cite{Osborne,Aubin-book}), an equilibrium for the whole system
is reached when every player's reaction is {}``unilaterally optimal'',
i.e., when, given the rival players' current strategies, any change
in a player's own strategy would result in a rate loss. This vector-valued
power control game was widely studied and several sufficient conditions
have been derived that guarantee the uniqueness of the NE and the
convergence of alternative distributed waterfilling based algorithms:
synchronous sequential \cite{Yu}-\cite{Scutari_Thesis}, synchronous
simultaneous \cite{Scutari_Thesis,Huang-Cendrillon,Sung-JSAC07,Scutari-Part II},
and asynchronous \cite{Scutari-ICASSP07,Scutari-IT-08}.

Interestingly, different approaches have been used in the cited papers
to analyze the game, most of them based on the following, apparently
different, key results: 1) the interpretation of the waterfilling
operator as a projection onto a proper polyhedral set \cite{Scutari_Thesis,Scutari-Part II};
2) the interpretation of the Nash equilibria of the game as solutions
of a proper affine Variational Inequality (VI) problem \cite{Luo-Pang};
and 3) the interpretation of the waterfilling mapping as a piecewise
affine function \cite[Ch. 4]{Sung-JSAC07,Facchinei}. In this paper,
we provide a unified view of these results, showing that they fit
naturally in our interpretation of the waterfilling mapping as a projector
\cite{Scutari_Thesis,Scutari-Part II}. Building on this interpretation
and using classical results from fixed-point and contraction theory
(cf. \cite{Bertsekas Book-Parallel-Comp,Ortega,Agarwal-book,Bellman}),
we then develop a mathematical framework useful to derive a unified
set of sufficient conditions guaranteeing both the uniqueness of the
NE and the convergence of totally asynchronous iterative waterfilling
based algorithms.

The proposed mathematical framework is instrumental to study the more
general MIMO case, which is a nontrivial extension of the SISO frequency-selective
case. There are indeed only a few papers that have studied (special
cases of) the MIMO game \cite{Scutari-GTMIMO}, \cite{Larsson-Jorswieck}-\cite{Arslan_etal}.
In \cite{Larsson-Jorswieck}, the authors focused on the two-user
MISO channel. In \cite{Ye-Blum-SP}-\cite{Liang-Dandekar-WC}, the
authors considered the rate maximization game in MIMO interference
channels, but they provided only numerical results to support the
existence of a NE of the game. Furthermore, in these papers there
is no study of the uniqueness of the equilibrium and convergence of
the proposed algorithms. Finally, in \cite{Arslan_etal}, the authors
showed that the MIMO rate maximization game is a concave game (in
the sense of \cite{Rosen}), implying the existence of a NE for any
set of arbitrary channel matrices \cite[Theorem 1]{Rosen}. As far
as the uniqueness of the equilibrium is concerned, they only showed
that if the multiuser interference is almost negligible, then the
NE is unique, without quantifying how small the interference must
be. Hence, a practical condition that one can check to guarantee the
uniqueness of the NE of the game and convergence of distributed algorithms
is currently missing.

The main difficulty in the MIMO case is that the optimal transmit
directions (i.e., eigenvectors of the transmit covariance matrix)
of each user change with the strategies of the other users, as opposed
to the SISO frequency-selective case where only the power allocation
depends on the strategies of the others, but the directions remain
fixed: i) in the diagonal MIMO case, the transmit directions are always
the canonical vectors \cite{Yu}-\cite{Scutari_Thesis}; ii) in the
frequency-selective channel, the transmit directions are the Fourier
vectors \cite{Scutari-Part I,Scutari-Part II}; iii) for the MISO
case, the transmit directions are matched to the vector channels;
and iv) for the SIMO case, there are no transmit directions to optimize.
For the previous reason, the existing results and techniques in \cite{Yu}-\cite{Scutari-IT-08},
valid for SISO frequency-selective channels, cannot be applied or
trivially extended to the MIMO case. On top of that, another difficulty
is the fact that, differently from the vector power control game in
\cite{Yu}-\cite{Scutari-Part II}, where there exists an explicit
relationship (via the waterfilling solution) among the optimal power
allocations of all the users, in the matrix-valued MIMO game one cannot
obtain an explicit expression of the optimal covariance matrix of
each user at the NE (the MIMO waterfilling solution), as a function
of the optimal covariance matrices of the other users, but there exists
only a complicated implicit relationship, via an eigedecomposition.

Building on the mathematical framework developed for the SISO case,
we can overcome the main difficulties in the study of the MIMO game
invoking a novel interpretation of the \emph{MIMO }waterfilling operator
as a projector and its nonexpansion property, similar to the one for
frequency-selective channels. This enables us to derive a unified
set of sufficient conditions that guarantee the uniqueness of the
Nash equilibrium of the MIMO game and the convergence of totally asynchronous
distributed algorithms based on the MIMO waterfilling solution. 

The paper is organized as follows. Section \ref{Sec:System.Model}
gives the system model and formulates the optimization problem as
a strategic noncooperative game. In Section \ref{Sec:Unified_Tool},
we draw the relationship between Nash equilibria of the game and fixed
points of nonlinear sets of equations, and provide the mathematical
tools necessary to study convergence of distributed asynchronous algorithms.
Building on the interpretation of the multiuser waterfilling solution
as a proper projection onto a convex set, in Section \ref{sec:WF-Contraction},
we provide the main properties of the multiuser waterfilling solution
either in the SISO or MIMO case, unifying previous results proposed
in the literature to study the rate maximization game in SISO frequency-selective
channels. The contraction property of the multiuser waterfilling paves
the way to derive sufficient conditions guaranteeing the uniqueness
of the fixed point of the waterfilling projector$-$alias the NE of
the (SISO/MIMO) game$-$and the convergence of iterative, possibly
asynchronous, distributed algorithms, as detailed in Sections \ref{Existence-Uniqueness}
and \ref{Section_Distributed Algorithms}, respectively. Section \ref{Num_Res}
reports some numerical results illustrating the benefits of MIMO transceivers
in the multiuser context. Finally, Section \ref{Sec:Conclusion} draws
some conclusions.

\section{System Model and Problem Formulation\label{Sec:System.Model}}

In this section we introduce the system model and formulate the optimization
problem addressed in the paper explicitly.

\subsection{System Model}

We consider a vector Gaussian interference channel composed of $Q$
links. In this model, there are $Q$ transmitter-receiver pairs, where
each transmitter wants to communicate with its corresponding receiver
over a MIMO channel. The transmission over the generic $q$-th MIMO
channel with $n_{T_{q}}$ transmit and $n_{R_{q}}$ receive dimensions
can be described with the baseband signal model\begin{equation}
\mathbf{y}_{q}=\mathbf{H}_{qq}\mathbf{x}_{q}+\sum_{r\neq q}\mathbf{H}_{rq}\mathbf{x}_{r}+\mathbf{n}_{q},\label{vector I/O}\end{equation}
where $\mathbf{x}_{q}\mathbf{\in\mathbb{C}}^{n_{T_{q}}}$ is the vector
transmitted by source $q$, $\mathbf{H}_{qq}\mathbf{\in\mathbb{C}}^{n_{R_{q}}\times n_{T_{q}}}$
is the direct channel of link $q$, $\mathbf{H}_{rq}\mathbf{\in\mathbb{C}}^{n_{R_{q}}\times n_{T_{r}}}$
is the cross-channel matrix between source $r$ and destination $q$,
$\mathbf{y}_{q}\mathbf{\in\mathbb{C}}^{n_{R_{q}}}$ is the vector
received by destination $q$, and $\mathbf{n}_{q}\mathbf{\in\mathbb{C}}^{n_{R_{q}}}$
is a zero-mean circularly symmetric complex Gaussian noise vector
with arbitrary covariance matrix $\mathbf{R}_{n_{q}}$(assumed to
be nonsingular). The second term on the right-hand side of (\ref{vector I/O})
represents the Multi-User Interference (MUI) received by the $q$-th
destination and caused by the other active links. For each transmitter
$q$, the total average transmit power is\begin{equation}
\mathcal{E}\left\{ \left\Vert \mathbf{x}_{q}\right\Vert _{2}^{2}\right\} =\mathsf{Tr}\left(\mathbf{Q}_{q}\right)\leq P_{q},\label{power-constraint}\end{equation}
where $\mathsf{Tr}\left(\cdot\right)$ denotes the trace operator,
$\mathbf{Q}_{q}\triangleq\mathcal{E}\left\{ \mathbf{x}_{q}\mathbf{x}_{q}^{H}\right\} $
is the covariance matrix of the transmitted vector $\mathbf{x}_{q}$,
and $P_{q}$ is the maximum average transmitted power in units of
energy per transmission.

The system model in (\ref{vector I/O})-(\ref{power-constraint})
provides a unified way to represent many physical communication channels
and multiuser systems of practical interest. What changes from one
system to the other is the structure of the channel matrices. We may
have, in fact, as particular cases of (\ref{vector I/O})-(\ref{power-constraint}):
i) digital subscriber lines \cite{Starr-Cioffi Book}, where the channel
matrices are Toeplitz circulant, the matrices $\mathbf{Q}_{q}=\mathbf{W}\limfunc{Diag}(\mathbf{p}_{q})\mathbf{W}^{H}$
incorporate the DFT precoding $\mathbf{W}^{H}$, the vectors $\mathbf{p}_{q}$
allocate the power across the frequency bins, and the MUI is mainly
caused by near-end cross talk; ii) single (or multi) antenna CDMA
cellular radio systems, where the matrices $\mathbf{Q}_{q}=\mathbf{F}_{q}\mathbf{F}_{q}^{H}$
contain in $\mathbf{F}_{q}$ the user codes within a given cell, and
the MUI is essentially intercell interference \cite{Goldsmith-Wicker};
iii) ad-hoc wireless MIMO networks, where the channel matrices represent
the MIMO channel of each link \cite{Akyildiz-Wang}. 

Since our goal is to find distributed algorithms that do not require
neither a centralized control nor a coordination among the links,
we focus on transmission techniques where no interference cancellation
is performed and multiuser interference is treated as additive colored
noise from each receiver. Each channel is assumed to change sufficiently
slowly to be considered fixed during the whole transmission, so that
the information theoretical results are meaningful. Moreover, perfect
channel state information at both transmitter and receiver sides of
each link is assumed;%
\footnote{Note that each user $q$ is only required to known his own channel
$\mathbf{H}_{qq}$, but not the cross-channels $\{\mathbf{H}_{rq}\}_{r\neq q}$
from the other users.%
} each receiver is also assumed to measure with no errors the covariance
matrix of the noise plus MUI generated by the other users. Finally,
we assume that the channel matrices $\mathbf{H}_{qq}$ are square
nonsingular. The more general case of possibly rectangular nonfull
rank matrices is addressed in \cite{Scutari-GTMIMO}.

Under these assumptions, invoking the capacity expression for the
single user Gaussian MIMO channel$-$achievable using random Gaussian
codes by all the users$-$the maximum information rate on link $q$
for a given set of users' covariance matrices $\mathbf{Q}_{1},\ldots,\mathbf{Q}_{Q}$
is \cite{Cover}\begin{equation}
R_{q}(\mathbf{Q}_{q},\mathbf{Q}_{-q})=\log\limfunc{det}\left(\mathbf{I}+\mathbf{H}_{qq}^{H}\mathbf{R}_{\mathbf{-}q}^{-1}(\mathbf{Q}_{-q})\mathbf{H}_{qq}\mathbf{Q}_{q}\right)\label{R_qq}\end{equation}
 where $\mathbf{R}_{\mathbf{-}q}(\mathbf{Q}_{-q})\triangleq\mathbf{R}_{n_{q}}+\sum\limits _{r\neq q}\mathbf{H}_{rq}\mathbf{Q}_{r}\mathbf{H}_{rq}^{H}$
is the MUI plus noise covariance matrix observed by user $q$, and
$\mathbf{Q}_{-q}\triangleq\left(\mathbf{Q}_{r}\right)_{r=1,\, r\neq q}^{Q}$
is the set of all the users' covariance matrices, except the $q$-th
one.

\subsection{Game Theoretical Formulation \label{Sec:GT formulation}}

We formulate the system design within the framework of game theory
using as desirable criterion the concept of Nash Equilibrium{\huge {}
}(NE) (cf. \cite{Osborne,Aubin-book}). Specifically, we consider
a strategic noncooperative game, in which the players are the links
and the payoff functions are the information rates on each link: Each
player $q$ competes against the others by choosing his transmit covariance
matrix $\mathbf{Q}_{q}$ (i.e., his strategy) that maximizes his own
information rate $R_{q}(\mathbf{Q}_{q},\mathbf{Q}_{-q})$ in (\ref{R_qq}),
subject to the transmit power constraint (\ref{power-constraint}).
A solution of the game$-$a NE$-$is reached when each user, given
the strategy profiles of the others,{\huge{} }does not get any rate
increase by unilaterally changing his own strategy. Stated in mathematical{\huge{}
}terms, the game has the following structure:\begin{equation}
\left(\mathscr{G}\right):\qquad\begin{array}{ll}
\limfunc{maximize}\limits _{\mathbf{Q}_{q}} & R_{q}(\mathbf{Q}_{q},\mathbf{Q}_{-q})\\
\limfunc{subject}\limfunc{to} & \mathbf{Q}_{q}\in{\mathscr{Q}}_{q},\end{array}\qquad\forall q\in\Omega,\qquad\qquad\qquad\label{Rate-matrix-game}\end{equation}
 where $\Omega\triangleq\{1,\ldots,Q\}$ is the set of players (i.e.,
the links); $R_{q}(\mathbf{Q}_{q},\mathbf{Q}_{-q})$ defined in (\ref{R_qq})
is the payoff function of player $q$; and ${\mathscr{Q}}_{q}$ is
the set of admissible strategies (the covariance matrices) for player
$q$, defined as%
\footnote{Observe that, in the definition of $\mathscr{Q}_{q}$ in (\ref{set_Q_q})
the condition $\mathbf{Q}=\mathbf{Q}^{H}$ is redundant, since any
\emph{complex} positive semidefinite matrix must be necessarily Hermitian
\cite[Sec. 7.1]{Horn85}. Furthermore, we replaced, without loss of
generality (w.l.o.g.), the original inequality power constraint in
(\ref{power-constraint}) with equality, since, at the optimum to
each problem in (\ref{Rate-matrix-game}), the constraint must be
satisfied with equality. %
} \begin{equation}
\mathscr{Q}_{q}\triangleq\left\{ \mathbf{Q}\in\mathcal{\ \mathbb{C}}^{n_{T_{q}}\times n_{T_{q}}}:\text{ }\mathbf{Q\succeq0},\quad\limfunc{Tr}\{\mathbf{Q}\}=P_{q}\right\} .\label{set_Q_q}\end{equation}

The solutions of game $\mathscr{G}$ are formally defined as follows.%
\footnote{Observe that, for the payoff functions defined in (\ref{R_qq}), we
can indeed limit ourselves to adopt pure strategies w.l.o.g., as we
did in (\ref{R_qq}), since every NE of the game can be proved to
be achievable using pure strategies \cite{Scutari-Part I}. %
}

\smallskip{}

\begin{definition} \label{NE def} \emph{A (pure) strategy profile}
$\mathbf{Q}^{\star}=\left(\mathbf{Q}_{q}^{\star}\right)_{q\in\Omega}\in{\mathscr{Q}}_{1}\times\ldots\times{\mathscr{Q}}_{Q}$
\emph{is a NE of game} ${\mathscr{G}}$ \emph{if} \begin{equation}
R_{q}(\mathbf{Q}_{q}^{\star},\mathbf{Q}_{-q}^{\star})\geq R_{q}(\mathbf{Q}_{q},\mathbf{Q}_{-q}^{\star}),\ \text{\ \ }\forall\mathbf{Q}_{q}\in{\mathscr{Q}}_{q},\text{ }\forall q\in\Omega.\label{pure-NE}\end{equation}
 \hspace{\fill}$\square$\end{definition}\smallskip{}

To write the Nash equilibria of game $\mathscr{G}$ in a convenient
form, we first introduce the following notations and definitions.
Given $\mathscr{G}$, for each $q\in\Omega$ and $\mathbf{Q}_{-q}\in\mathscr{Q}_{-q}\triangleq{\mathscr{Q}}_{1}\times\ldots\times{\mathscr{Q}}_{q-1},{\mathscr{Q}}_{q+1}\times\ldots\times{\mathscr{Q}}_{Q}$,
we write the eigendecomposition of $\mathbf{H}_{qq}^{H}\mathbf{R}_{-q}^{-1}(\mathbf{Q}_{-q})\mathbf{H}_{qq}$
as: \begin{equation}
\mathbf{H}_{qq}^{H}\mathbf{R}_{-q}^{-1}(\mathbf{Q}_{-q})\mathbf{H}_{qq}\mathbf{=}\mathbf{U}_{q}\mathbf{D}_{q}\mathbf{U}_{q}^{H},\label{eq:multiuser_eigendecomposition}\end{equation}
 where $\mathbf{U}_{q}=\mathbf{U}_{q}(\mathbf{Q}_{-q})\in\mathbf{\mathbb{C}}^{n_{T_{q}}\times n_{T_{q}}}$
is a unitary matrix with the eigenvectors, $\mathbf{D}_{q}=\mathbf{D}_{q}(\mathbf{Q}_{-q})\in\mathbb{R}_{++}^{n_{T_{q}}\times n_{T_{q}}}$
is a diagonal matrix with the $n_{T_{q}}$ positive eigenvalues, and
$\mathbf{R}_{-q}(\mathbf{Q}_{-q})=\mathbf{R}_{n_{q}}+\sum\limits _{r\neq q\hfill}\mathbf{H}_{rq}\mathbf{Q}_{r}\mathbf{H}_{rq}^{H}$.

Given $q\in\Omega$ and $\mathbf{Q}_{-q}\in\mathscr{Q}_{-q}$, the
solution to problem (\ref{Rate-matrix-game}) is the well-known waterfilling
solution (e.g., \cite{Cover}): \begin{equation}
\begin{array}{c}
\mathbf{Q}_{q}^{\star}=\mathsf{\mathbf{WF}}_{q}(\mathbf{Q}_{-q}^{}),\end{array}\label{Best_response_WF-single-user}\end{equation}
with the waterfilling operator $\mathbf{WF}_{q}\left(\mathbf{\cdot}\right)$
defined as \begin{equation}
\mathbf{WF}_{q}\left(\mathbf{Q}_{-q}\right)\triangleq\mathbf{U}_{q}\left(\mu_{q}\mathbf{I-D}_{q}^{-1}\right)^{+}\mathbf{U}_{q}^{H},\label{WF_MIMO_op}\end{equation}
where $\mathbf{U}_{q}=\mathbf{U}_{q}(\mathbf{Q}_{-q})$, $\mathbf{D}_{q}=\mathbf{D}_{q}(\mathbf{Q}_{-q})$,
and $\mu_{q}$ is chosen to satisfy $\limfunc{Tr}\left\{ (\mu_{q}\mathbf{I-D}_{q}^{-1})^{+}\right\} =P_{q}$,
with $(x)^{+}\triangleq\max(0,x).$

Using (\ref{Best_response_WF-single-user}) and Definition \ref{NE def},
we can now characterize the Nash Equilibria of the game $\mathscr{G}$
in a compact way as the following waterfilling fixed-point equation:
\begin{equation}
\begin{array}{c}
\mathbf{Q}_{q}^{\star}=\mathsf{\mathbf{WF}}_{q}(\mathbf{Q}_{-q}^{\star})\end{array},\quad\forall q\in\Omega.\label{Best_response_WF}\end{equation}

\bigskip{}
\noindent\setcounter{rem}{1}\emph{Remark \therem{} - Competitive
maximization of transmission rates.} The choice of the objective function
as in (\ref{R_qq}) requires the use of ideal Gaussian codebooks with
a proper covariance matrix. In practice, Gaussian codes may be substituted
with simple (suboptimal) finite-order signal constellations, such
as Quadrature Amplitude Modulation (QAM) or Pulse Amplitude Modulation
(PAM), and practical (suboptimal) coding schemes. In this case, instead
of considering the maximization of \emph{mutual information} on each
link, one can focus on the competitive maximization of the \emph{transmission
rate}, using finite order constellation, under constraints on transmit
power and on the average error probability $P_{e,q}^{\star}$ (see
\cite{Scutari-Part I} for more details). Interestingly, using a similar
approach to that in \cite{Scutari-Part I}, one can prove that the
optimal transmission strategy of each user is still a solution to
the fixed-point equation in (\ref{Best_response_WF}), where each
channel matrix $\mathbf{H}_{qq}$ is replaced by $\mathbf{H}_{qq}/\Gamma_{q}$,
where $\Gamma_{q}\geq1$ denotes the gap, which depends only on the
constellations and on the target error probability $P_{e,q}^{\star}$
\cite{Forney-91 }. \medskip{}

\addtocounter{rem}{1}\noindent\emph{Remark \therem{} - Related
works}\textbf{\emph{.}} The matrix nature of game $\mathscr{G}$ and
the arbitrary structure of the channel matrices make the analysis
of the game quite complicated, since none of the results in game theory
literature can be directly applied to characterize solutions of the
form (\ref{Best_response_WF}). The main difficulty in the analysis
comes from the fact that, for each user $q,$ the optimal eigenvector
matrix $\mathbf{U}_{q}^{\star}=\mathbf{U}_{q}(\mathbf{Q}_{-q}^{\star})$
in (\ref{WF_MIMO_op}) depends, in general, on the strategies $\mathbf{Q}_{-q}^{\star}$
of all the others, through a very complicated implicit relationship$-$the
eigendecomposition of the equivalent channel matrix $\mathbf{H}_{qq}^{H}\mathbf{R}_{-q}^{-1}(\mathbf{Q}_{q}^{\star})\mathbf{H}_{qq}.$

\noindent In the vector power control games studied in \cite{Yu}-\cite{Scutari-IT-08},
the analysis of uniqueness of the equilibrium was mathematically more
tractable, since scalar frequency-selective channels are represented
by diagonal matrices (or Toeplitz and circulant matrices \cite{Scutari-Part I,Scutari-Part II}),
implying that the optimal set of eigenvectors of any NE becomes \emph{user-independent}
\cite{Scutari-Part I}. In the present case, it follows that, because
of the dependence of the optimal strategy $\mathbf{U}_{q}^{\star}(\mathbf{Q}_{-q}^{\star})$
of each user on the strategy profile of the others at the NE, one
cannot use the uniqueness condition of the NE obtained in \cite{Yu}-\cite{Scutari-IT-08}
to guarantee the uniqueness of the NE of game $\mathscr{G}$ in (\ref{Rate-matrix-game}),
even if game $\mathscr{G}$ reduces to the power control game studied
in the cited papers, once the optimal users' strategy profile in (\ref{Best_response_WF})
is introduced in (\ref{Rate-matrix-game}). At the best of our knowledge,
the only paper where game $\mathscr{G}$ was partially analyzed is
\cite{Arslan_etal}, where the authors applied the framework developed
in \cite{Rosen} to the MIMO game and showed that the NE becomes unique
if the MUI in the system$-$the interference-to-noise ratio at each
receiver$-$is sufficiently small, but without quantifying exactly
how much small the MUI must be. Thus, a practical condition that one
can check to guarantee the uniqueness of the NE is still missing.\medskip{}

To overcome the difficulties in the study of game $\mathscr{G}$,
we propose next an equivalent expression of the waterfilling solution
enabling us to express the Nash equilibria in (\ref{Best_response_WF})
as a fixed-point of a more tractable mapping. This alternative expression
is based on the new interpretation of the MIMO waterfilling solution
as a proper projector operator. Based on this result, we can then
derive sufficient conditions for the uniqueness of the NE and convergence
of asynchronous distributed algorithms, as detailed in Sections \ref{Existence-Uniqueness}
and \ref{Section_Distributed Algorithms} respectively.

\section{Nash Equilibrium as a Fixed Point\label{Sec:Unified_Tool}}

Before providing one of the major results of the paper$-$the contraction
properties of the MIMO multiuser waterfilling projector$-$we recall
and unify some standard results from fixed-point \cite{Agarwal-book}
and contraction theory \cite{Bertsekas Book-Parallel-Comp,Ortega}
that will be instrumental for our derivations (recall from (\ref{Best_response_WF})
that any NE can be interpreted as a fixed point of the waterfilling
mapping). The proposed unified mathematical framework is also useful
to establish an interesting  link among the alternative, apparently
different, approaches proposed in the literature to study the rate
maximization game in SISO frequency-selective interference channels
\cite{Yu}-\cite{Scutari_Thesis}, showing that most of the results
in \cite{Yu}-\cite{Scutari_Thesis} can be unified by our interpretation
of the waterfilling as a projector \cite{Scutari_Thesis,Scutari-Part II}.

\subsection{Existence and Uniqueness of a Fixed-point\label{Sec_ex_uniq}}

Let $\mathbf{T}:\mathcal{X}\mapsto\mathcal{X}$ be any mapping from
a subset $\mathcal{X}\subseteq\mathbb{R}^{n}$ to itself. One can
associate $\mathbf{T}$ to a dynamical system described by the following
discrete-time equation:\begin{equation}
\mathbf{x(}n+1\mathbf{)=T(x}(n)\mathbf{),\quad}n\in\mathbb{N}_{+},\label{dynamical-system}\end{equation}
where $\mathbf{x}(n)\in\mathbb{R}^{n}$ is the vector of the state
variables of the system at (discrete) time $n,$ with $\mathbf{x}(0)=\mathbf{x}_{0}\in\mathbb{R}^{n}.$
The equilibria of the system, if they exist, are the vectors $\mathbf{x}^{\star}$
resulting as a solution of $\mathbf{x}^{\star}=\mathbf{T(x}^{\star}),$
i.e., the fixed-points of mapping $\mathbf{T}$. The study of the
existence and uniqueness of an equilibrium of a dynamical system has
been widely addressed either in fixed-point theory (cf. \cite{Bertsekas Book-Parallel-Comp,Ortega,Agarwal-book})
or control theory (cf. \cite{Bellman,Khalil,Liberzon-book})  literature.
Many alternative conditions are available. Throughout this paper,
we will use the following.\smallskip{}

\begin{theorem} \label{Theo_existemce-uniq}\emph{Given the dynamical
system in (\ref{dynamical-system}) with} $\mathbf{T}:\mathcal{X}\mapsto\mathcal{X}$
and $\mathcal{X}\subseteq\mathbb{R}^{n},$ \emph{we have the following:}

\noindent a) Existence\emph{ }(\emph{cf.} \cite{Ortega,Agarwal-book}):
\emph{If} $\mathcal{X}$ \emph{is nonempty, convex and compact, and}
$\mathbf{T}$ \emph{is a continuous mapping, then there exists some}
$\mathbf{x}^{\star}$ \emph{such that} $\mathbf{x}^{\star}=\mathbf{T(x}^{\star});$

\noindent b) Uniqueness (\emph{cf. }\cite{Bertsekas Book-Parallel-Comp,Ortega,Agarwal-book})\emph{:}
\emph{If} $\mathcal{X}$ \emph{is closed and }$\mathbf{T}$ \emph{is
a contraction in some vector norm} $\left\Vert \mathbf{\cdot}\right\Vert ,$
\emph{with modulus} $\alpha\in\lbrack0,1),$ \emph{i.e.}, \begin{equation}
\left\Vert \mathbf{T}(\mathbf{x}^{(1)})-\mathbf{T}(\mathbf{x}^{(2)})\right\Vert \leq\alpha\left\Vert \mathbf{x}^{(1)}-\mathbf{x}^{(2)}\right\Vert ,\quad\forall\mathbf{x}^{(1)},\mathbf{x}^{(2)}\in\mathcal{X}{,}\label{contraction}\end{equation}
\emph{then the fixed-point of} $\mathbf{T}$ \emph{is unique}. \hfill $\square$\end{theorem}\smallskip{}

\noindent\addtocounter{rem}{1}\emph{Remark \therem}\textbf{\emph{
- }}\emph{Sufficiency of the conditions.} The conditions of Theorem
\ref{Theo_existemce-uniq} are only sufficient for the existence and
uniqueness of the fixed point. However, this does not mean that some
of them can be removed. For example, the convexity assumption in the
existence condition cannot, in general, be removed, as the simple
one-dimensional example $T(x)=-x$ and $\mathcal{X}=\left\{ -c,c\right\} ,$
with$\ c\in\mathbb{R},$ shows.\medskip{}

\noindent\addtocounter{rem}{1}\emph{Remark \therem}\textbf{\emph{
- }}\emph{Choice of the norm.} The contractive property of the mapping
is \textit{norm-dependent}, in the sense that a mapping may be contractive
for some choice of the norm on \textbf{\ }$\mathbb{R}^{n}$ and,
at the same time, it may fail to be so under a different norm. On
the other hand, it may happen that a mapping is a contraction in more
than one norm. In such a case, even though the uniqueness of the fixed-point
is guaranteed whatever the choice of the norm is (cf. Theorem \ref{Theo_existemce-uniq}),
the convergence of different algorithms, based on the same mapping
$\mathbf{T}$, to the fixed-point is, in general, norm-dependent.
Thus, the choice of the proper norm is a critical issue and it actually
gives us potential degrees of freedom to be explored in the characterization
of the convergence properties of the desired algorithms used to reach
the fixed-point. We address this issue in the next sections, where
we introduce a proper norm, tailored to our needs.

\subsection{Convergence to a Fixed-point\label{Sec:fixed_point_convergence}}

\noindent Nonlinear fixed-point problems are typically solved by iterative
methods, especially when one is interested in distributed algorithms
\cite{Bertsekas Book-Parallel-Comp,Ortega}. In fact, the mapping
$\mathbf{x}\mapsto\mathbf{T(x)}$ can be interpreted as an algorithm
for finding such a fixed point. The degrees of freedom are in the
choice of the specific updating scheme among the components of vector
$\mathbf{x}$, based on mapping $\mathbf{T}$. More specifically,
denoting by $\mathbf{x=(x}_{1},\ldots,\mathbf{x}_{Q})$ a partition
of $\mathbf{x}$, with $\mathbf{x}_{q}\in\mathbb{R}^{n_{q}}$ and
$n_{1}+\ldots+n_{Q}=n,$ and assuming $\mathcal{X=X}_{1}\times\cdots\times\mathcal{X}_{Q},$%
\footnote{For the sake of simplicity, we focus on mappings $\mathbf{T}$ whose
domain can be written as the Cartesian product of lower dimensional
sets, associated to the partition of the mapping. For our purposes,
this choice is enough, since the joint admissible strategy set of
game $\mathscr{G}$ satisfies this condition.%
} with each $\mathcal{X}_{q}\subseteq\mathbb{R}^{n_{q}},$ the most
common updating strategies for $\mathbf{x}_{1},\ldots,\mathbf{x}_{Q}$
based on mapping $\mathbf{T}$ are \cite{Bertsekas Book-Parallel-Comp,Ortega}:

\begin{description}
\item [{\textmd{i)}}] \emph{Jacobi scheme: }All components $\mathbf{x}_{1},\ldots,\mathbf{x}_{Q}$
are updated \emph{simultaneously,} via the mapping $\mathbf{T}$;
\item [{\textmd{ii)}}] \emph{Gauss-Seidel scheme:} All components $\mathbf{x}_{1},\ldots,\mathbf{x}_{Q}$
are updated \emph{sequentially,} one after the other, via the mapping
$\mathbf{T}$;
\item [{\textmd{iii)}}] \emph{Totally asynchronous scheme:} All components
$\mathbf{x}_{1},\ldots,\mathbf{x}_{Q}$ are updated in a \emph{totally
asynchronous way} (in the sense of \cite{Bertsekas Book-Parallel-Comp}),
via the mapping $\mathbf{T.}$ According to this scheme, some components
$\mathbf{x}_{q}$ may be updated more frequently than others and,
when they are updated, a possibly \emph{outdated} information on the
other components can be used. Some variations of such a totally asynchronous
scheme, e.g., including constraints on the maximum tolerable delay
in the updating and on the use of the outdated information (which
leads to the so-called \emph{partially} asynchronous algorithms),
can also be considered \cite{Bertsekas Book-Parallel-Comp}. 
\end{description}
Observe that the latter algorithm contains, as special cases, the
first two ones. In general, the above algorithms converge to the fixed-point
of $\mathbf{T}$ under different conditions \cite{Bertsekas Book-Parallel-Comp,Ortega}.
However, we can obtain a unified set of convergence conditions (not
necessarily the mildest ones) by studying the contraction properties
of mapping $\mathbf{T}$ under a proper choice of the norm. To prove
the convergence of the totally asynchronous algorithms, a useful norm
is the so-called \emph{block-maximum} norm, defined as follows. According
to the partition $\mathbf{x}_{1},\ldots,\mathbf{x}_{Q}$ of $\mathbf{x}$
and $\mathbf{T=(T}_{q}\mathbf{)}_{q=1}^{Q}$, with $\mathbf{T}_{q}:\mathcal{X}\mapsto\mathcal{X}_{q},$
let $\left\Vert \mathbf{\cdot}\right\Vert _{q}$ denote any vector
norm on $\mathbb{R}^{n_{q}}$ for each $q,$ the block-maximum norm
on $\mathbb{R}^{n}$ is defined as \cite{Bertsekas Book-Parallel-Comp,Ortega}
\begin{equation}
\left\Vert \mathbf{T}\right\Vert _{\text{block}}=\max_{q}\left\Vert \mathbf{T}_{q}\right\Vert _{q}.\label{block-maximum}\end{equation}
The mapping $\mathbf{T}$ is said to be a block-contraction with modulus
$\alpha\in\lbrack0,1)$ if it is a contraction in the block-maximum
norm with modulus $\alpha$. A unified set of convergence conditions
for distributed algorithms based on mapping $\mathbf{T}$ is given
in the following theorem, whose proof follows the same steps as in
\cite[Appendix A]{Scutari-IT-08} and is omitted here because of the
space limitation (see also \cite{Scutari-GTMIMO}).\smallskip{}

\begin{theorem} \label{unified-set-of-conditions}\emph{Given the
mapping} $\mathbf{T=(T}_{q}\mathbf{)}_{q=1}^{Q}:\mathcal{X}\mapsto\mathcal{X}$,
with $\mathcal{X=X}_{1}\times\cdots\times\mathcal{X}_{Q},$ \emph{assume
that} $\mathbf{T}$\emph{ is a block-contraction with modulus }$\alpha\in\lbrack0,1).$
\emph{Then, the totally asynchronous algorithm }(\emph{cf.} \cite{Bertsekas Book-Parallel-Comp})\emph{
based on the mapping }$\mathbf{T}$ \emph{asymptotically converges
to the unique fixed-point of} $\mathbf{T}$, \emph{for any set of
initial conditions in} $\mathcal{X}$ \emph{and updating schedule}.
\hfill $\square$ \end{theorem}\smallskip{}

Theorem \ref{unified-set-of-conditions} provides a unified set of
convergence conditions for all the algorithms that are special cases
of the totally asynchronous algorithm. Weaker conditions can still
be obtained if one is not interested in a totally asynchronous implementation.
For example, if only the Jacobi updating scheme is considered, to
prove the contraction property of $\mathbf{T}$, one can use \emph{any}
arbitrary norm on $\mathbb{R}^{n}$ \cite{Bertsekas Book-Parallel-Comp,Ortega}.

\subsection{Contraction Theory, Lyapunov Function, and Variational Inequality
Problems }

Contraction theory is not the only instrument available to prove the
convergence of distributed algorithms to the fixed-point of a mapping
$\mathbf{T}$. So far, we have seen that any mapping $\mathbf{T}$
defines a dynamical system (see (\ref{dynamical-system})). Hence,
the convergence of distributed algorithms to the fixed-point of $\mathbf{T}$
can be reformulated as the study of the globally asymptotic stability
of the equilibrium of a proper dynamical system, based on $\mathbf{T.}$
From this perspective, Lyapunov theory is a valuable instrument to
study the system behavior and, as a by-product, the convergence of
distributed algorithms \cite{Bellman,Khalil}. Indeed, the contraction
property of mapping $\mathbf{T}$ in the vector norm $\left\Vert \cdot\right\Vert $
implies the existence of a valid Lyapunov function for the dynamical
system in (\ref{dynamical-system}) \cite{Bellman,Khalil}, given
by $V(\mathbf{x})=\left\Vert \mathbf{x}\mathbf{-x^{\star}}\right\Vert .$
This guarantees the convergence of a Jacobi scheme based on mapping
$\mathbf{T}$. Interestingly, in the case of contraction of mapping
$\mathbf{T}$ in the block maximum norm (\ref{block-maximum}), the
Lyapunov function $V(\mathbf{x})=\left\Vert \mathbf{x}\mathbf{-x^{\star}}\right\Vert _{\text{block}}$
can be interpreted as the \emph{common} Lyapunov function of a set
of interconnected dynamical systems, each of them associated to the
partition $\mathbf{x}_{1},\ldots,\mathbf{x}_{Q}$ of $\mathbf{x}$
\cite{Liberzon-book}.

\noindent \indent Finally, it is interesting to observe that the
convergence to the fixed-point of a mapping $\mathbf{T}$ can also
be studied introducing a proper transformation of $\mathbf{T}$ that
preserves the set of the fixed-points. A useful tool to explore this
direction is given by the variational inequality theory \cite{CPStone92,Facchinei}
(see, e.g., \cite{Luo-Pang,Scutari-Part II} and Section \ref{WF_as_AVI}
for an application of this framework to the multiuser waterfilling
mapping).

We are now ready to apply the previous general framework to the multiuser
waterfilling mapping in (\ref{WF_MIMO_op}), as detailed next.

\section{Contraction Properties of the Multiuser Waterfilling Mapping \label{sec:WF-Contraction}}

So far we have seen that a unified set of conditions guaranteeing
the uniqueness of the NE and the convergence of totally asynchronous
algorithms to the fixed-point Nash equilibria of game $\mathscr{G}$
can be obtained deriving conditions for the multiuser waterfilling
mapping in (\ref{WF_MIMO_op}) to be a contraction in a proper block-maximum
norm (see Theorems \ref{Theo_existemce-uniq} and \ref{unified-set-of-conditions}).
In this section we then provide a contraction theorem for the multiuser
waterfilling operator. Our result is based on the interpretation of
MIMO waterfilling operator as a matrix projection onto the convex
set of feasible strategies of the users. This result is also useful
to obtain a unified view of, apparently different, techniques used
in the literature to study the uniqueness of the NE and the convergence
of alternative waterfilling based algorithms in the rate-maximization
game over \emph{SISO frequency-selective} Gaussian interference channels
\cite{Yu}-\cite{Scutari-IT-08}. To show this interesting relationship,
we start from an overview of the main properties of the Nash equilibria
of game $\mathscr{G}$ in the case of SISO frequency-selective\emph{
}Gaussian interference channels, as obtained in \cite{Scutari_Thesis,Scutari-Part II,Scutari-IT-08},
\cite{Luo-Pang}, \cite{Sung-JSAC07}, and \cite{Yu,Huang-Cendrillon},
and then we consider the more general MIMO case.\vspace{-0.2cm}

\subsection{Frequency-Selective Gaussian Interference Channels \label{Sec:Frequency-Selective-case}\label{WF_as_AVI}}

In the case of block transmission over SISO frequency-selective channels
\cite{Yu}-\cite{Scutari-IT-08}, each channel matrix $\mathbf{H}_{rq}\in\mathbb{C}^{N\times N}$
becomes a Toeplitz circulant matrix and $n_{T_{q}}=n_{R_{q}}=N,$
where $N$ is the length of the transmitted block (see, e.g., \cite{Scutari-Part I}).
This leads to the eigendecomposition $\mathbf{H}_{rq}$ $=\mathbf{WD}_{rq}\mathbf{W}^{H}$,
where $\mathbf{W\in\mathbb{C}}^{N\times N}$ is the normalized IFFT
matrix, i.e., $\left[\mathbf{W}\right]_{ij}\triangleq e^{j2\pi(i-1)(j-1)/N}/\sqrt{N}$
for $i,j=1,\ldots,N$ and $\mathbf{D}_{rq}$ is a $N\times N$ diagonal
matrix, where $\left[\mathbf{D}_{rq}\right]_{kk}\triangleq H_{rq}(k)$
is the frequency-response of the channel between source $r$ and destination
$q$ at carrier $k$, and $\mathbf{R}_{n_{q}}=\limfunc{Diag}(\sigma_{q}^{2}(1),\ldots,\sigma_{q}^{2}(N))$.
Under this setup, the matrix game $\mathscr{G}$ in (\ref{Rate-matrix-game})
reduces to a simpler power control game, where the strategy of each
user $q$ becomes the power allocation $\mathbf{p}_{q}=[p_{q}(1),\ldots,p_{q}(N)]^{T}$
over the $N$ carriers and the admissible strategy set in (\ref{set_Q_q})
reduces to%
\footnote{In the transmissions over frequency-selective channels, in addition
to the power constraints as in (\ref{set_P_q}), it may be useful
to introduce spectral mask constraints, in order to impose radiation
limits over some (licensed) bands \cite{Luo-Pang,Scutari-Part II,Scutari-IT-08,Scutari_Thesis}.
Interestingly, all the results presented on this section are valid
also in the presence of spectral mask constraints, as proved in \cite{Luo-Pang,Scutari-Part II,Scutari-IT-08,Scutari_Thesis}.%
}\begin{equation}
\mathscr{P}_{q}\triangleq\left\{ \mathbf{x}\in\mathcal{\ \mathbb{R}}_{+}^{N}:\sum_{k=1}^{N}x_{k}=P_{q},\quad\forall k\in\{1,\ldots,N\}\right\} ,\label{set_P_q}\end{equation}
and $\mathscr{P}\triangleq{\mathscr{P}}_{1}\times\ldots\times{\mathscr{P}}_{Q}$.
It follows that the optimal strategy at any NE must satisfy the simultaneous
multiuser waterfilling equation: \begin{equation}
\begin{array}{c}
\mathbf{p}_{q}^{\star}=\mathsf{\mathbf{wf}}_{q}(\mathbf{p}_{-q}^{\star}),\end{array}\quad\forall q\in\Omega,\label{OFDM-Waterfilling}\end{equation}
where $\mathbf{p}_{-q}\triangleq\left(\mathbf{p}_{r}\right)_{r\in\Omega,r\neq q}$,
and the waterfilling operator $\mathsf{\mathbf{wf}}_{q}\left(\mathbf{\cdot}\right)$
becomes \cite{Scutari-Part II} \begin{equation}
\left[\mathsf{\mathbf{wf}}_{q}\left(\mathbf{p}_{-q}\right)\right]_{k}\triangleq\left(\mu_{q}-\dfrac{\sigma_{q}^{2}(k)+\sum_{\, r\neq q}|H_{rq}(k)|^{2}p_{r}(k)}{|H_{qq}(k)|^{2}}\right)^{+},\label{WF_OFDM}\end{equation}
for $k\in\{1,\ldots,N\},$ with the waterlevel $\mu_{q}$ chosen to
satisfy the power constraint $\sum_{k=1}^{N}p_{q}(k)=P_{q}$.

Different approaches have been proposed in the literature to study
the properties of the Nash equilibria in (\ref{OFDM-Waterfilling}),
each time obtaining milder conditions for the uniqueness and the convergence
of distributed algorithms \cite{Yu}-\cite{Scutari-IT-08}. We provide
in the following a unified view of the techniques used in the cited
papers, based on the mathematical framework described in Section \ref{Sec:Unified_Tool}.
\bigskip{}

\noindent\textbf{Approach \#1: Multiuser waterfilling as a projector
\cite{Scutari_Thesis,Scutari-Part II}} \textbf{-} We introduce first
the following intermediate definitions. For any given $\mathbf{p}_{-q}$,
let $\boldsymbol{\mathsf{insr}}_{q}(\mathbf{p}_{-q})$ be the $N$-dimensional
vector defined as\begin{equation}
\left[\boldsymbol{\mathsf{insr}}_{q}(\mathbf{p}_{-q})\right]_{k}\triangleq\dfrac{\sigma_{q}^{2}(k)+\sum_{\, r\neq q}|H_{rq}(k)|^{2}p_{r}(k)}{|H_{qq}(k)|^{2}},\label{INR_q}\end{equation}
with $k\in\{1,\ldots,N\}$. In order to apply Theorem \ref{unified-set-of-conditions},
we introduce a proper block-maximum norm for the multiuser waterfilling
mapping $\mathbf{wf}=(\mathbf{wf}_{q}(\mathbf{p}_{-q}))_{q\in\Omega}$
in (\ref{WF_OFDM}) (cf. Section \ref{Sec:fixed_point_convergence}).
Given some $\mathbf{w}\triangleq[w_{q},\ldots,w_{Q}]^{T}>\mathbf{0}$,
let $\left\Vert \cdot\right\Vert _{2,\text{block}}^{\mathbf{w}}$
denote the (vector) block-maximum norm$,$ defined as \cite{Bertsekas Book-Parallel-Comp}
\begin{equation}
\left\Vert \mathbf{wf}(\mathbf{p})\right\Vert _{2,\text{block}}^{\mathbf{w}}\triangleq\max_{q\in\Omega}\frac{\left\Vert \mathbf{wf}_{_{q}}(\mathbf{p}_{-q})\right\Vert _{2}}{w_{q}},\text{ }\quad\label{block_max_weight_norm-vector}\end{equation}
 where $\left\Vert \mathbf{\cdot}\right\Vert _{2}$ is the Euclidean
norm. Let $\left\Vert \mathbf{\cdot}\right\Vert _{\infty,\text{vec}}^{\mathbf{w}}$
be the \emph{vector} weighted maximum norm$,$ defined as \cite{Horn85}
\begin{equation}
\left\Vert \mathbf{x}\right\Vert _{\infty,\text{vec}}^{\mathbf{w}}\triangleq\max_{q\in\Omega}\frac{\left\vert x_{q}\right\vert }{w_{q}},\quad\mathbf{w>0,}\text{ }\quad\mathbf{x\in\mathbb{R}}^{Q},\label{weighted_infinity_vector_norm}\end{equation}
and let $\left\Vert \mathbf{\cdot}\right\Vert _{\infty,\text{mat}}^{\mathbf{w}}$
denote the \emph{matrix} norm induced by $\left\Vert \cdot\right\Vert _{\infty,\text{vec}}^{\mathbf{w}},$
given by \cite{Horn85} \begin{equation}
\left\Vert \mathbf{A}\right\Vert _{\infty,\text{mat}}^{\mathbf{w}}\triangleq\max_{q}\frac{1}{w_{q}}\dsum\limits _{r=1}^{Q}\left\vert [\mathbf{A}]_{qr}\right\vert \text{ }w_{r},\text{ }\quad\mathbf{A\in\mathbb{R}}^{Q\times Q}.\label{H_max_weight_norm}\end{equation}
We also introduce the nonnegative matrix $\mathbf{S}^{\max}\in\mathbb{R}_{+}^{Q\times Q}$,
defined as \begin{equation}
\left[\mathbf{S}^{\max}\right]_{qr}\triangleq\left\{ \begin{array}{ll}
\max\limits _{k\in\{1,\ldots N\}}\dfrac{|H_{qr}(k)|^{2}}{|H_{qq}(k)|^{2}}, & \text{if }\ r\neq q,\\
0, & \text{otherwise.}\end{array}\right.\label{Hmax}\end{equation}

\noindent Using the above definitions and denoting by $\left[\mathbf{x}_{0}\right]_{\mathscr{X}}=\limfunc{argmin}_{\mathbf{z}\in\mathscr{X}}\left\Vert \mathbf{z}\mathbf{-}\mathbf{x}_{0}\right\Vert _{2}$
the Euclidean projection of vector $\mathbf{x}_{0}$ onto the convex
set $\mathscr{X}$, in \cite{Scutari_Thesis,Scutari-Part II} we proved
the following. \smallskip{}

\begin{lemma}[\emph{Waterfilling as a projector}] \label{Lemma_wf_OFDM}\emph{The
waterfilling operator} ${\mathbf{wf}}_{q}\left(\mathbf{p}_{-q}\right)$
\emph{in} (\ref{WF_OFDM}) \emph{can be equivalently written as}\begin{equation}
\mathsf{wf}_{q}\left(\mathbf{p}_{-q}\right)=\left[-\boldsymbol{\mathsf{insr}}_{q}(\mathbf{p}_{-q})\right]_{{\mathscr{P}}_{q}},\label{WF-projection}\end{equation}
\emph{where} $\mathscr{P}_{q}$ \emph{and} $\boldsymbol{\mathsf{insr}}_{q}(\cdot)$
\emph{are defined in} (\ref{set_P_q}) \emph{and} (\ref{INR_q}),
\emph{respectively}. \hfill $\square$\end{lemma}\smallskip{}

\indent It follows from Lemma \ref{Lemma_wf_OFDM} that the Nash
equilibria in (\ref{OFDM-Waterfilling}) can be alternatively obtained
as the fixed-points of the mapping defined in (\ref{WF-projection})\begin{equation}
\mathbf{p}_{q}^{\star}=\left[-\boldsymbol{\mathsf{insr}}_{q}(\mathbf{p}_{-q}^{\star})\right]_{{\mathscr{P}}_{q}},\quad\forall q\in\Omega.\label{eq:fixed-point_NE_SISO}\end{equation}
Lemma \ref{Lemma_wf_OFDM} is also the key result to study contraction
properties of mapping $\mathbf{wf}$ and thus, based on (\ref{eq:fixed-point_NE_SISO}),
to derive conditions for the uniqueness of the NE and convergence
of distributed algorithms. The main result is summarized in the following
theorem that comes from \cite[Proposition 2]{Scutari-Part II}.%
\footnote{We refer the interested reader to \cite{Scutari-Part II,Scutari-IT-08}
for mode general results on contraction properties of the $\mathbf{wf}$
mapping.%
}

\smallskip{}

\begin{theorem}[\emph{Contraction property of the mapping} $\mathbf{wf}$]
\label{Theorem_contraction_SISO}\emph{Given} $\mathbf{w}\triangleq[w_{1},\ldots,w_{Q}]^{T}\mathbf{>0}$,
\emph{the mapping} $\mathbf{wf}$ \emph{defined in} (\ref{WF_OFDM})
\emph{is} Lipschitz continuous\emph{ on $\mathscr{P}$}:\emph{ }\begin{equation}
\begin{array}{lc}
\left\Vert \mathbf{wf(p}^{(1)}\mathbf{)}-\mathbf{wf(p}^{(2)}\mathbf{)}\right\Vert _{2,\mathrm{block}}^{\mathbf{w}}\leq & \Vert\mathbf{S}^{\max}\Vert_{\infty,\text{mat}}^{\mathbf{w}}\quad\quad\quad\quad\quad\quad\\
 & \times\left\Vert \mathbf{p}^{(1)}-\mathbf{p}^{(2)}\right\Vert _{2,\mathrm{block}}^{\mathbf{w}},\quad\quad\end{array}\label{b_modulus_contraction}\end{equation}
$\forall\mathbf{p}^{(1)},\text{ }\mathbf{p}^{(2)}\in{\mathscr{P}}\text{ },$
\emph{where} $\left\Vert \cdot\right\Vert _{2,\text{block}}^{\mathbf{w}},$
$\left\Vert \cdot\right\Vert _{\infty,\text{mat}}^{\mathbf{w}}$,
\emph{and} $\mathbf{S}^{\max}$ \emph{are defined in} (\ref{block_max_weight_norm-vector}),
(\ref{H_max_weight_norm}), \emph{and} (\ref{Hmax}), \emph{respectively}.
\emph{Furthermore}, \emph{if} \begin{equation}
\Vert\mathbf{S}^{\max}\Vert_{\infty,\text{mat}}^{\mathbf{w}}<1\mathbf{,}\label{eq:SF-block-maximum-norm}\end{equation}
 \emph{then mapping} $\mathbf{wf\ }$ \emph{is a} \textit{\emph{block-contraction
with modulus}} $\Vert\mathbf{S}^{\max}\Vert_{\infty,\text{mat}}^{\mathbf{w}}.$
\hfill $\square$\end{theorem}\smallskip{}

Given Theorem \ref{Theorem_contraction_SISO}, it follows from Theorem
\ref{Theo_existemce-uniq} and Theorem \ref{unified-set-of-conditions},
that condition $\Vert\mathbf{S}^{\max}\Vert_{\infty,\text{mat}}^{\mathbf{w}}<1$
is sufficient to guarantee the uniqueness of the NE of the game as
well as the convergence of totally asynchronous algorithm based on
the waterfilling mapping $\mathbf{wf}$ in (\ref{WF_OFDM}) \cite{Scutari-Part II,Scutari-IT-08}. 

\bigskip{}

\noindent\textbf{Approach \#2: Multiuser waterfilling as solution
of an Affine VI \cite{Luo-Pang} - }In \cite{Luo-Pang}, the authors
established an interesting reformulation of the rate maximization
game as a linear complementarity problem (LCP) \cite{CPStone92}.
More specifically, they proved that the nonlinear system of KKT optimality
conditions of the $q$-th user convex problem in $\mathscr{G}$, given
by {[}where $a\perp b$ means that the two scalars $a$ and $b$ are
orthogonal, i.e., $a\cdot b=0$]%
\footnote{Observe that in \cite{Luo-Pang}, the authors considered equivalently
for each user $q\in\Omega$ the power constraint $\sum_{k=1}^{N}p_{q}(k)\leq P_{q},$
rather than $\sum_{k=1}^{N}p_{q}(k)=P_{q}$, as we did in (\ref{set_P_q}).%
}\begin{equation}
\begin{array}{rc}
-\dfrac{|H_{qq}(k)|^{2}}{\sigma_{q}^{2}(k)+\sum_{r=1}^{Q}|H_{rq}(k)|^{2}p_{r}(k)}+\mu_{q}-\nu_{q,k}=0, & \quad\forall k,\\
0\leq\mu_{q}\perp\left(P_{q}-\sum_{k=1}^{N}p_{q}(k)\right)\geq0,\\
0\leq\nu_{q,k}\perp p_{q}(k)\geq0, & \quad\forall k,\end{array}\label{eq:KKT_rate_max_row-Luo-Pang}\end{equation}
is equivalent to \cite[Proposition 1]{Luo-Pang}:\begin{equation}
\begin{array}{r}
\hspace{-0.3cm}0\leq p_{q}(k)\perp\left(\dfrac{\sigma_{q}^{2}(k)}{|H_{qq}(k)|^{2}}+\sum\limits _{r=1}^{Q}\dfrac{|H_{rq}(k)|^{2}}{|H_{qq}(k)|^{2}}p_{r}(k)+\lambda_{q}\right)\geq0,\\
\forall k,\\
\lambda_{q}=\mbox{free, }\quad\sum_{k=1}^{N}p_{q}(k)=P_{q}.\end{array}\label{eq:KKT_rate_max_row1-Luo-Pang-system}\end{equation}
As observed in \cite{Luo-Pang}, (\ref{eq:KKT_rate_max_row1-Luo-Pang-system})
for all $q\in\Omega$ represents the KKT conditions of the Affine
VI (AVI) ($\mathscr{P}$, $\tilde{\boldsymbol{\sigma}},$$\mathbf{M}$)
defined by the polyhedral set $\mathscr{P}$ and the affine mapping
$\mathbf{p}\mapsto$ $\boldsymbol{\tilde{\sigma}}+\mathbf{M}\mathbf{p}$
(see \cite{CPStone92} for more details on the AVI problems), where
$\mathbf{p}\triangleq[\mathbf{p}_{1}^{T},\ldots,\mathbf{p}_{Q}^{T}]^{T}$,
$\boldsymbol{\tilde{\sigma}}\triangleq[\boldsymbol{\boldsymbol{\tilde{\sigma}}}_{1}^{T},\ldots,\boldsymbol{\boldsymbol{\tilde{\sigma}}}_{Q}^{T}]^{T}$,
and $\mathbf{M}$ is a block partition matrix $\mathbf{M}\triangleq\left[\mathbf{M}_{rq}\right]_{r,q\in\Omega}$,
with $\mathbf{p}_{q}=[p_{q}(1),\ldots,p_{q}(N)]^{T},$ $\boldsymbol{\tilde{\sigma}}_{q}=[\sigma_{q}^{2}(1)/|H_{qq}(1)|^{2},\ldots,\sigma_{q}^{2}(N)$
$/|H_{qq}(N)|^{2}]^{T}$ and $\mathbf{M}_{qr}\triangleq\limfunc{Diag}\left(|H_{rq}(1)|^{2}/|H_{qq}(1)|^{2},\ldots,|H_{rq}(N)|^{2}/|H_{qq}(N)|^{2}\right)$,
for $r,q\in\Omega$.

It follows that the vector $\mathbf{p}^{\star}\in\mathscr{P}$ is
a NE of the game $\mathscr{G}$ if and only if it satisfies the AVI
($\mathscr{P}$, $\tilde{\boldsymbol{\sigma}},$$\mathbf{M}$) \cite{CPStone92,Facchinei}:\begin{equation}
(\mathbf{p}-\mathbf{p}^{\star})^{T}\left(\boldsymbol{\tilde{\sigma}}+\mathbf{M}\mathbf{\mathbf{p}^{\star}}\right)\geq0,\quad\forall\mathbf{p}\in\mathscr{P}.\label{eq:AVI}\end{equation}

Building on this result and the properties of AVI problems (cf. \cite{CPStone92,Facchinei}),
the authors in \cite{Luo-Pang} derived sufficient conditions for
the uniqueness of the Nash equilibrium and the global convergence
of synchronous sequential IWFA. 

It can be show that the AVI ($\mathscr{P}$, $\boldsymbol{\tilde{\sigma}}$,
$\mathbf{M}$) in (\ref{eq:AVI}) is equivalent to the fixed-point
equation in (\ref{eq:fixed-point_NE_SISO}) \cite{Facchinei,CPStone92},
establishing the link between the solutions to (\ref{eq:AVI})$-$the
fixed-points of the waterfilling mapping $\mathbf{wf}$ in (\ref{WF_OFDM})$-$and
the interpretation of the mapping $\mathbf{wf}$ as a projection (Lemma
\ref{Lemma_wf_OFDM}), as given in \cite{Scutari_Thesis,Scutari-Part II}\footnote{Papers \cite{Scutari-Part I, Scutari-Part II} have been originally submitted on September 2004.}. In fact, the convergence conditions obtained
in \cite{Luo-Pang} for the synchronous sequential IWFA coincide with
(\ref{eq:SF-block-maximum-norm}) (for a proper choice of vector $\mathbf{w}$
\cite{Scutari-Part II}).%
\footnote{Recall that condition (\ref{eq:SF-block-maximum-norm}) is valid also
for the convergence of the \emph{asynchronous} IWFA (Theorem \ref{unified-set-of-conditions})
\cite{Scutari-IT-08}.%
} Observe that they are a special case of those obtained in \cite[Corollary 2]{Scutari-IT-08}. 

\bigskip{}

\noindent\textbf{Approach \#3: Multiuser waterfilling as a piecewise
affine function \cite[Ch. 4]{Sung-JSAC07,Facchinei} - }In \cite{Sung-JSAC07},
the authors proved global convergence of different waterfilling based
algorithms building on the key result that the waterfilling mapping
$\mathbf{wf}:\mathscr{P}\mapsto\mathscr{P}$ can be equivalently written
as a piecewise affine (PA) function on $\mathbb{R}^{QN}$ \cite[Ch. 4]{Facchinei}.
In fact, the result in \cite{Sung-JSAC07} follows the same steps
as Propositions 4.1.1 and 4.2.2 in \cite[Ch. 4]{Facchinei} and, interestingly,
can be obtained directly from our interpretation of the $\mathbf{wf}$
as a projection (Lemma \ref{Lemma_wf_OFDM}) and some properties of
the PA functions in \cite[Ch. 4]{Facchinei}, as detailed next. We
introduce the following intermediate definitions first.\smallskip{}

\noindent\emph{Definition 2} (\cite[Def. 4.1.3]{Facchinei}): \emph{A
continuous function }\textbf{\emph{$\mathbf{f}:\mathbb{R}^{n}\mapsto\mathbb{R}^{m}$}}
\emph{is said to be }piecewise affine\emph{ (PA) if there exists a
finite family of affine functions $\{\mathbf{f}_{k}(\mathbf{x})=\mathbf{A}_{k}\mathbf{x}+\mathbf{b}_{k}\}_{k=1}^{K}$
for some positive integer $K$ and $\{(\mathbf{A}_{k},\mathbf{b}_{k})\}_{k=1}^{K}$,
with each }$\mathbf{f}_{k}:\mathbb{R}^{n}\mapsto\mathbb{R}^{m}$,
\emph{such that for all }$\mathbf{x}\in\mathbb{R}^{n}$, $\mathbf{f}(\mathbf{x})\in\{\mathbf{f}_{1}(\mathbf{x}),\ldots,\mathbf{f}_{K}(\mathbf{x})\}$.\hfill $\square$\smallskip{}

PA functions have many interesting properties (we refer the interested
reader to \cite[Ch. 4]{Facchinei} for an in-depth study of the theory
of PA functions). Here, we are interested in the following one, which
follows directly from \cite[Proposition 4.2.2 (c)]{Facchinei}. \smallskip{}

\begin{lemma}\label{Lemma_Contraction_PA} \emph{Any PA map }\textbf{\emph{$\mathbf{f}:\mathbb{R}^{n}\mapsto\mathbb{R}^{m}$}}
\emph{is }globally Lipschitz continuous\emph{ on }$\mathbb{R}^{n}$:\emph{
}\begin{eqnarray}
\left\Vert \mathbf{f(x}^{(1)}\mathbf{)}-\mathbf{f(x}^{(2)}\mathbf{)}\right\Vert _{\text{vec}} & \leq & \alpha\left\Vert \mathbf{x}^{(1)}-\mathbf{x}^{(2)}\right\Vert _{\text{vec}},\nonumber \\
 &  & \quad\forall\mathbf{x}^{(1)},\text{ }\mathbf{x}^{(2)}\in\mathbb{R}^{n}\mathbf{,}\label{PA_contraction}\end{eqnarray}
\emph{with Lipschitz constant }$\alpha\triangleq\max_{k\in\{1,\ldots,K\}}\Vert\mathbf{A}_{k}\Vert_{\text{mat}}$,
\emph{where} $\left\Vert \cdot\right\Vert _{\text{vec}}$ \emph{is
any vector norm and} $\left\Vert \cdot\right\Vert _{\text{mat}}$
\emph{is the matrix norm induced by} $\left\Vert \cdot\right\Vert _{\text{vec}}$
\emph{.}\hfill $\square$ \end{lemma}\smallskip{}

The link between our interpretation of the $\mathbf{wf}$ mapping
as a projector (Lemma \ref{Lemma_wf_OFDM}) and the interpretation
of $\mathbf{wf}$ as PA map (\cite[Theorem 5]{Sung-JSAC07}) is given
by the following \cite[Prop. 4.1.4]{Facchinei}.\smallskip{}

\begin{lemma}\emph{Let $\mathscr{X}$ be a polyhedral set in $\mathbb{R}^{n}$.}%
\footnote{Roughly speaking, a polyhedral is the intersection of a finite number
of halfspaces and hyperplanes (see, e.g., \cite[Ch. 2.2.4]{Boyd-book}).%
}\emph{ Then, the Euclidean projector onto $\mathscr{X}$ is a PA function
on $\mathbb{R}^{n}$.}\hfill $\square$\end{lemma} \smallskip{}

According to Lemma \ref{Lemma_wf_OFDM}, for any given $\mathbf{p}\geq\mathbf{0}$,
the waterfilling $\mathbf{wf}(\mathbf{p})$ in (\ref{WF_OFDM}) is
the Euclidean projector of vector $-\boldsymbol{\mathsf{insr}}(\mathbf{p})\triangleq-[\boldsymbol{\mathsf{insr}}_{1}(\mathbf{p}_{-1})^{T},\ldots,\boldsymbol{\mathsf{insr}}_{Q}(\mathbf{p}_{-Q})^{T}]^{T}$
onto the convex set $\mathscr{P}$, which is a \emph{polyhedral set.
}It then follows from Lemma 3 that $\mathbf{wf}(\mathbf{p})$ is a
PA function, i.e., there exists a finite family of affine functions
$\{\mathbf{f}_{k}(\mathbf{p})=\mathbf{A}_{k}\mathbf{p}+\mathbf{b}_{k}\}_{k=1}^{K}$
such that, for every $\mathbf{p\geq0}$ we have \begin{equation}
\mathbf{wf}(\mathbf{p})=\mathbf{A}_{k}\mathbf{p}+\mathbf{b}_{k},\qquad\mbox{for some}\qquad k\in\{1,\ldots K\},\label{eq:WF_as_PA}\end{equation}
which coincides with the result in \cite[Theorem 5]{Sung-JSAC07}.
The expression of the affine pieces $\{(\mathbf{A}_{k},\mathbf{b}_{k})\}_{k=1}^{K}$
of $\mathbf{wf}(\mathbf{p})$ can be obtained exploring the structure
of the waterfilling solution (\ref{WF_OFDM}). We omit the details
here because of space limitations (see (\cite[Theorem 5]{Sung-JSAC07}). 

Contraction properties of the $\mathbf{wf}$ operator interpreted
as PA map on $\mathbb{R}^{QN}$ as stated in \cite[Theorem 7]{Sung-JSAC07}
are a direct consequence of properties of the PA functions. In particular,
it follows from Lemma \ref{Lemma_Contraction_PA} that, if \begin{equation}
\alpha=\max_{k\in\{1,\ldots,K\}}\Vert\mathbf{A}_{k}\Vert_{\text{mat}}<1,\label{eq:SF-Uniqueness-NE-PA}\end{equation}
then the $\mathbf{wf}$ mapping is a contraction in the vector norm
$\Vert\cdot\Vert_{\text{vec}}$. Exploring different vector norms,
one can easily obtain different sufficient conditions for the uniqueness
of the NE of the game and the convergence of distributed algorithms,
based on the $\mathbf{wf}$ mapping. Observe that, according to Theorem
\ref{unified-set-of-conditions}, if a block-maximum norm is used
in (\ref{PA_contraction}) (see also Theorem \ref{Theorem_contraction_SISO}
and Theorem \ref{Theorem_contraction} in Section \ref{Sec:Contraction-MIMO}),
then condition (\ref{eq:SF-Uniqueness-NE-PA}) guarantees also the
global convergence of \emph{totally asynchronous} algorithms \cite{Scutari-IT-08,Sung-JSAC07}.

\bigskip{}

\noindent\textbf{Approach \#4: Multiuser waterfilling via the max-lemma
\cite{Yu,Huang-Cendrillon} - }In \cite{Huang-Cendrillon}, the authors,
among all, derived conditions for the global convergence of the synchronous
sequential and simultaneous iterative waterfilling algorithms (implying
also the uniqueness of the NE). Conditions in \cite{Huang-Cendrillon}
generalize those obtained in \cite{Yu} for the two-player game to
the case of arbitrary number of players. We show now that results
in \cite{Huang-Cendrillon} come from Theorem \ref{unified-set-of-conditions}
as special case, if a proper vector norm is chosen. To this end, we
introduce first the following intermediate definitions and results.

A key point in the proof of convergence in \cite{Huang-Cendrillon}
is given by the following max-lemma \cite[Lemma 1]{Huang-Cendrillon}.\smallskip{}

\begin{lemma} \emph{\label{Lemma_Janwei_et_al}} Let $f:\mathbb{R}\mapsto\mathbb{R}$
and $g:\mathbb{R}\mapsto\mathbb{R}$ \emph{be a non-decreasing and
non-increasing functions on} $\mathbb{R}$, \emph{respectively}. \emph{If
there exists a unique} $x^{\star}$ \emph{such that} $f(x^{\star})=g(x^{\star}),$
\emph{and function} $f$ \emph{and} $g$ \emph{are strictly increasing
and strictly decreasing at} $x=x^{\star},$ \emph{then}\begin{equation}
x^{\star}=\limfunc{argmin}_{x}\{\max\{f(x),g(x)\}\}.\label{x_unique}\end{equation}
 \hfill $\square$\end{lemma}

Let us introduce the following error dynamic, as defined in \cite{Huang-Cendrillon}:\begin{equation}
\begin{array}{r}
e_{q}(n+1)\triangleq\max\left\{ \dsum\limits _{k}\left[p_{q}^{(n+1)}(k)-p_{q}^{(n)}(k)\right]^{+},\:\:\:\right.\\
\left.\dsum\limits _{k}\left[p_{q}^{(n+1)}(k)-p_{q}^{(n)}(k)\right]^{-}\right\} ,\end{array}\label{error_scalar}\end{equation}
with $n\in\mathbb{N}_{+}=\left\{ 0,1,2,\ldots\right\} ,$ where $\mathbf{p}_{q}^{(n)}=[p_{q}^{(n)}(1),\ldots,p_{q}^{(n)}(N)]^{T}$
is the vector of the power allocation of user $q\in\Omega$, generated
at the discrete time $n$ by the sequential or simultaneous IWFA {[}starting
from any arbitrary feasible point $\mathbf{p}^{(0)}$], $\lbrack x]^{+}\triangleq\max(0,x)$,
and $\lbrack x]^{-}=\max(0,-x).$ Using Lemma \ref{Lemma_Janwei_et_al}
and following the same steps as in \cite{Huang-Cendrillon}, results
in \cite{Huang-Cendrillon} can be restated in terms of the error
vector $\mathbf{e}^{(n+1)}\triangleq[e_{1}^{(n+1)},\ldots,e_{Q}^{(n+1)}]$
as: \begin{equation}
\left\Vert \mathbf{e}^{(n+1)}\right\Vert _{\infty,\text{vec}}\leq\alpha\left\Vert \mathbf{e}^{(n)}\right\Vert _{\infty,\text{vec}},\quad\forall n\in\mathbb{N}_{+},\label{eq:error_contraction}\end{equation}
where $\alpha\triangleq(Q-1)\max_{k,\, r\neq q}\dfrac{|H_{rq}(k)|^{2}}{|H_{qq}(k)|^{2}}$,
and $\left\Vert \cdot\right\Vert _{\infty,\text{vec}}$ denotes the
$l_{\infty}$ norm {[}see (\ref{weighted_infinity_vector_norm}) with
$\mathbf{w}=\mathbf{1}$]. It follows from (\ref{eq:error_contraction})
that, under $\alpha<1$, both synchronous sequential and simultaneous
IWFAs globally converge to the unique NE of the game \cite{Yu,Huang-Cendrillon}.
Observe that this condition implies the contraction of the waterfilling
mapping as given in Theorem \ref{Theorem_contraction_SISO}, showing
the more generality of our sufficient condition (\ref{eq:SF-block-maximum-norm})
than that in \cite{Yu,Huang-Cendrillon}.

\smallskip{}

\noindent\textbf{Extension}: We generalize now the results in \cite{Yu,Huang-Cendrillon},
so that we can use Theorem \ref{unified-set-of-conditions} and enlarge
the convergence conditions of \cite{Yu,Huang-Cendrillon}, making
them to coincide with (\ref{eq:SF-block-maximum-norm}) and valid
also for the asynchronous IWFA \cite{Scutari-IT-08}. To this end,
we introduce a new vector norm, as detailed next. 

Inspired by (\ref{error_scalar}), we  introduce the following norm:\begin{equation}
\left\Vert \mathbf{x}\right\Vert _{1,\infty,\text{vec}}\triangleq\max\left\{ \left\Vert (\mathbf{x})^{+}\right\Vert _{1,\text{vec}},\left\Vert (\mathbf{x})^{-}\right\Vert _{1,\text{vec}}\right\} ,\qquad\mathbf{x}\in\mathbb{R}^{N},\label{eq:new-norm}\end{equation}
where $\left\Vert \cdot\right\Vert _{1,\text{vec}}$ denotes the $l_{1}$
norm \cite{Horn85}. Some properties of $\left\Vert \cdot\right\Vert _{1,\infty,\text{vec}}$
are listed in the following lemma (we omit the proof because of space
limitation).

\smallskip{}

\begin{lemma}\label{Lemma-alternative-nonexpansion-property}\emph{The
norm} $\left\Vert \cdot\right\Vert _{1,\infty,\text{vec}}$ in (\ref{eq:new-norm})
\emph{is a valid vector norm} (\emph{in the sense that it satisfies
the axioms of a norm} \cite{Horn85}). \emph{Moreover, the following
nonexpansion property holds true}:%
\footnote{Interestingly, one can prove that the nonexpansion property as stated
in (\ref{eq:nonexpansion-new-norm}) also holds true if in (\ref{eq:nonexpansion-new-norm})
the norm $\left\Vert \cdot\right\Vert _{1,\infty,\text{vec}}$ is
replaced by the $l_{1}$ norm.%
}\begin{eqnarray}
\hspace{-0.1cm}\left\Vert \left(\mu_{x}\mathbf{1}-\mathbf{x}_{0}\right)^{+}-\left(\mu_{y}\mathbf{1}-\mathbf{y}_{0}\right)^{+}\right\Vert _{1,\infty,\text{vec}} & \hspace{-0.1cm}\leq\hspace{-0.1cm} & \left\Vert \mathbf{x}_{0}-\mathbf{y}_{0}\right\Vert _{1,\infty,\text{vec}},\nonumber \\
 &  & \hspace{-0.1cm}\forall\mathbf{x}_{0},\mathbf{y}_{0}\in\mathbb{R}_{+}^{N},\label{eq:nonexpansion-new-norm}\end{eqnarray}
\emph{where the waterlevels} $\mu_{x}$ \emph{and} $\mu_{y}$ \emph{satisfy}
$\mathbf{1}^{T}$ $\left(\mu_{x}\mathbf{1}-\mathbf{x}_{0}\right)^{+}=\mathbf{1}^{T}\left(\mu_{y}\mathbf{1}-\mathbf{y}_{0}\right)^{+}=P_{T}$,
(\emph{with} $P_{T}$ \emph{an arbitrary nonnegative number}), \emph{and}
$\mathbf{1}$ \emph{denotes} \emph{the} $N$-\emph{dimensional vector
of all ones}.\hfill $\square$\end{lemma}

\smallskip{}

Interestingly, Lemma \ref{Lemma-alternative-nonexpansion-property}
provides the nonexpansion property of the (single-user) waterfilling
solution in the vector norm $\left\Vert \cdot\right\Vert _{1,\infty,\text{vec}}$.
It also represents the key result to prove the contraction property
of the multiuser waterfilling mapping $\mathbf{wf}$ in (\ref{WF_OFDM})
in the block-maximum norm $\left\Vert \cdot\right\Vert _{1,\infty,\text{block}}^{\mathbf{w}}$
associated to $\left\Vert \cdot\right\Vert _{1,\infty,\text{vec}}$,
defined, for each $\mathbf{w}\triangleq[w_{1},\ldots,w_{Q}]^{T}\mathbf{>0}$,
as \begin{equation}
\left\Vert \mathbf{wf}(\mathbf{p})\right\Vert _{1,\infty,\text{block}}^{\mathbf{w}}\triangleq\max_{q\in\Omega}\frac{\left\Vert \mathbf{wf}_{_{q}}(\mathbf{p}_{-q})\right\Vert _{1,\infty,\text{vec}}}{w_{q}}.\text{ }\quad\label{block_max_weight_norm-vector-strange-norm}\end{equation}
 The main result is stated next (the proof is based on Lemma \ref{Lemma-alternative-nonexpansion-property}
and follows similar steps of that in \cite[Appendix A]{Scutari-IT-08};
see also Theorem \ref{Theorem_contraction} in Section \ref{Sec:Contraction-MIMO}).
\smallskip{}

\begin{theorem}[\emph{Contraction property of mapping $\mathbf{wf}$}]
\label{Theorem_contraction_SISO-alternative-norm}\emph{Given} $\mathbf{w}\triangleq[w_{1},\ldots,w_{Q}]^{T}\mathbf{>0}$,
\emph{the mapping} $\mathbf{wf}$ \emph{defined in} (\ref{WF_OFDM})
\emph{is} Lipschitz continuous\emph{ on $\mathscr{P}$}: \begin{equation}
\begin{array}{lc}
\hspace{-0.2cm}\left\Vert \mathbf{wf(p}^{(1)}\mathbf{)}-\mathbf{wf(p}^{(2)}\mathbf{)}\right\Vert _{1,\infty,\mathrm{block}}^{\mathbf{w}}\leq\hspace{-0.2cm} & \Vert\mathbf{S}^{\max}\Vert_{\infty,\text{mat}}^{\mathbf{w}}\quad\quad\quad\quad\quad\quad\\
 & \hspace{-0.2cm}\times\left\Vert \mathbf{p}^{(1)}-\mathbf{p}^{(2)}\right\Vert _{1,\infty,\mathrm{block}}^{\mathbf{w}},\quad\quad\end{array}\label{b_modulus_contraction-new-norm}\end{equation}
\emph{$\forall\mathbf{p}^{(1)},\text{ }\mathbf{p}^{(2)}\in{\mathscr{P}}$,
where} $\left\Vert \cdot\right\Vert _{1,\infty,\text{block}}^{\mathbf{w}}$,
$\left\Vert \cdot\right\Vert _{\infty,\text{mat}}^{\mathbf{w}}$,
\emph{and }$\mathbf{S}^{\max}$ \emph{are defined in} (\ref{block_max_weight_norm-vector-strange-norm}),
(\ref{H_max_weight_norm}), \emph{and} (\ref{Hmax}), \emph{respectively}.
\hfill $\square$\end{theorem}\smallskip{}

Comparing Theorem \ref{Theorem_contraction_SISO} with Theorem \ref{Theorem_contraction_SISO-alternative-norm},
one infers that both theorems provide the same sufficient conditions
for the waterfilling mapping $\mathbf{wf}$ to be a contraction and
thus the same sufficient conditions guaranteeing the uniqueness of
the NE and the convergence of asynchronous IWFAs \cite{Scutari-Part II,Scutari-IT-08}.

\subsection{MIMO Gaussian Interference Channels }

In this section, we generalize our interpretation of the waterfilling
projector in the frequency-selective case to the MIMO multiuser case.
For the sake of simplicity, we concentrate on MIMO systems whose direct
channel matrices $\mathbf{H}_{qq}$ are square and nonsingular. The
more general case is much more involved and goes beyond the scope
of the present paper; it has been considered in \cite{Scutari-GTMIMO}.\medskip{}

\subsubsection{Multiuser waterfilling in Gaussian MIMO interference channels \label{Sec:WF-MIMO-proj}}

We first introduce the following intermediate result. \smallskip{}

\begin{proposition}\label{Prop_MIMO_WF_eq}\emph{Given} $\mathbf{R}_{n_{}}\succ\mathbf{0}$,
$\mathbf{H}\in\mathbb{C}^{n\times n}$, \emph{and} $P_{T}>0$, \emph{let
define the following two convex optimization problems}: \begin{equation}
\mbox{\emph{(P1)}}:\qquad\begin{array}{ll}
\limfunc{maximize}\limits _{\mathbf{X}\succeq\mathbf{0}} & \:\:\,\log\det\left(\mathbf{R}_{n_{}}+\mathbf{H}\mathbf{XH}^{H}\right)\\
\limfunc{subject}\limfunc{to} & \begin{array}{l}
\limfunc{Tr}\{\mathbf{X}\}\leq P_{T},\end{array}\end{array}\label{single-user-capacity}\end{equation}
\emph{and} 

\begin{equation}
\mbox{\emph{(P2)}}:\qquad\begin{array}{ll}
\limfunc{minimize}\limits _{\mathbf{X}\succeq\mathbf{0}} & \left\Vert \mathbf{X-X}_{0}\right\Vert _{F}^{2}\smallskip\\
\limfunc{subject}\text{ }\limfunc{to} & \limfunc{Tr}\{\mathbf{X}\}=P_{T}.\end{array}\qquad\quad\quad\label{convex-projection}\end{equation}
\emph{If} $\mathbf{X}_{0}$ \emph{in} (P2) \emph{is chosen as} $\mathbf{X}_{0}=-\left(\mathbf{H}^{H}\mathbf{R}_{n}^{-1}\mathbf{H}\right)^{-1}$,
\emph{then both problems} (P1) \emph{and }(P2)\emph{ have the same
unique solution}.\end{proposition}

\begin{proof}Problem (P1) {[}and (P2)] is convex and admits a unique
solution, since the objective function is strictly concave (and strictly
convex) on $\mathbf{X}\succeq\mathbf{0}$. The Lagrangian function
$\mathcal{L}$ associated to (\ref{single-user-capacity}) is\begin{equation}
\mathcal{L}=-\log\det\left(\mathbf{R}_{n}+\mathbf{H}\mathbf{XH}^{H}\right)-\limfunc{Tr}(\boldsymbol{\Psi}\mathbf{X})+\lambda\left(\limfunc{Tr}(\mathbf{X})-P_{T}\right),\label{eq:Lagrangian}\end{equation}
which leads to the following KKT optimality conditions (Slater's conditions
are satisfied \cite[Ch. 5.9.1]{Boyd-book}):\vspace{-0.3cm}

\begin{eqnarray}
-\mathbf{H}^{H}\left(\mathbf{R}_{n}+\mathbf{H}\mathbf{XH}^{H}\right)^{-1}\mathbf{H}+\lambda\mathbf{I} & = & \boldsymbol{\Psi},\label{eq:KKT_rate_max_row1}\\
\boldsymbol{\Psi}\succeq\mathbf{0},\quad\mathbf{X}\succeq\mathbf{0},\quad\limfunc{Tr}(\boldsymbol{\Psi}\mathbf{X}) & = & 0,\label{eq:KKT_rate_max_row2}\\
\lambda\geq0,\quad\lambda\left(\limfunc{Tr}(\mathbf{X})-P_{T}\right)=0,\quad\limfunc{Tr}(\mathbf{X}) & \leq & P_{T}.\label{eq:KKT_rate_max_row3}\end{eqnarray}
First of all, observe that $\lambda$ must be positive. Otherwise,
(\ref{eq:KKT_rate_max_row1}) would lead to\begin{equation}
\mathbf{0}\succ-\mathbf{H}^{H}\left(\mathbf{R}_{n}+\mathbf{H}\mathbf{XH}^{H}\right)^{-1}\mathbf{H}=\boldsymbol{\Psi}\succeq\mathbf{0},\label{eq:contraddiction}\end{equation}
which cannot be true. We rewrite now (\ref{eq:KKT_rate_max_row1})-(\ref{eq:KKT_rate_max_row3})
in a more convenient form. To this end, we introduce \begin{equation}
\mathbf{X}_{0}\triangleq-\left(\mathbf{H}^{H}\mathbf{R}_{n}^{-1}\mathbf{H}\right)^{-1}\prec\mathbf{0},\label{eq:X_0}\end{equation}
so that\begin{equation}
\mathbf{H}^{H}\left(\mathbf{R}_{n}+\mathbf{H}\mathbf{XH}^{H}\right)^{-1}\mathbf{H}=\left(\mathbf{X}-\mathbf{X}_{0}\right)^{-1}\succ\mathbf{0}.\label{eq:X_minus_X_zero}\end{equation}
Then, using  the fact that $\lambda>0$ and absorbing
in (\ref{eq:KKT_rate_max_row1})-(\ref{eq:KKT_rate_max_row2}) the slack variable $\boldsymbol{\Psi}$,
system (\ref{eq:KKT_rate_max_row1})-(\ref{eq:KKT_rate_max_row3})
can be rewritten as
%
\begin{eqnarray}
\mathbf{X}\left[-\left(\mathbf{X}-\mathbf{X}_{0}\right)^{-1}+\lambda\mathbf{I}\right] & = & \mathbf{0},\label{eq:KKT_rate_max_row1_v3}\\
\boldsymbol{X\succeq\mathbf{0},}\quad-\left(\mathbf{X}-\mathbf{X}_{0}\right)^{-1}+\lambda\mathbf{I} & \succeq & \mathbf{0},\label{eq:KKT_rate_max_row2_v3}\\
\lambda>0,\quad\limfunc{Tr}(\mathbf{X}) & = & P_{T},\label{eq:KKT_rate_max_row3_v3}\end{eqnarray}
where in (\ref{eq:KKT_rate_max_row1_v3}) we have used the following
fact \cite[fact 8.10.3]{Bernstein} \begin{equation}
\limfunc{Tr}(\boldsymbol{\Psi}\mathbf{X})=0\quad\Leftrightarrow\quad\boldsymbol{\Psi}\mathbf{X}=\mathbf{0},\quad\quad\quad\forall\boldsymbol{\Psi},\mathbf{X}\succeq\mathbf{0}.\label{eq:Trace-equivalence}\end{equation}
Since $\lambda>0$, (\ref{eq:KKT_rate_max_row1_v3})-(\ref{eq:KKT_rate_max_row3_v3})
become

\begin{eqnarray}
\mathbf{X}\left[\left(\mathbf{X}-\mathbf{X}_{0}\right)-\frac{1}{\lambda}\mathbf{I}\right] & = & \mathbf{0},\label{eq:KKT_rate_max_row1_v4}\\
\boldsymbol{X\succeq\mathbf{0},}\quad-\left(\mathbf{X}-\mathbf{X}_{0}\right)^{-1}+\lambda\mathbf{I} & \succeq & \mathbf{0},\label{eq:KKT_rate_max_row2_v4}\\
\lambda>0,\quad\limfunc{Tr}(\mathbf{X}) & = & P_{T}.\label{eq:KKT_rate_max_row3_v4}\end{eqnarray}
We show now that (\ref{eq:KKT_rate_max_row1_v4})-(\ref{eq:KKT_rate_max_row3_v4})
is equivalent to 

\begin{eqnarray}
\mathbf{X}\left[\left(\mathbf{X}-\mathbf{X}_{0}\right)+\mu\mathbf{I}\right] & = & \mathbf{0},\label{eq:KKT_proj_row1_v2}\\
\boldsymbol{X}\succeq\mathbf{0},\quad\mathbf{X}-\mathbf{X}_{0}+\mu\mathbf{I} & \succeq & \mathbf{0},\label{eq:KKT_proj_row2_v2}\\
\mu=\mbox{free},\quad\limfunc{Tr}(\mathbf{X}) & = & P_{T}.\label{eq:KKT_proj_row3_v2}\end{eqnarray}
(\ref{eq:KKT_rate_max_row1_v4})-(\ref{eq:KKT_rate_max_row3_v4})
$\Rightarrow$ (\ref{eq:KKT_proj_row1_v2})-(\ref{eq:KKT_proj_row3_v2}):
Let $(\mathbf{X},\lambda)$ be a solution of (\ref{eq:KKT_rate_max_row1_v4})-(\ref{eq:KKT_rate_max_row3_v4}).
A solution of (\ref{eq:KKT_proj_row1_v2})-(\ref{eq:KKT_proj_row3_v2})
is obtained using $(\mathbf{X},\mu)$, with $\mu=-\frac{1}{\lambda}$.\smallskip{}

\noindent(\ref{eq:KKT_proj_row1_v2})-(\ref{eq:KKT_proj_row3_v2})
$\Rightarrow$ (\ref{eq:KKT_rate_max_row1_v4})-(\ref{eq:KKT_rate_max_row3_v4}):
Let $(\mathbf{X},\mu)$ be a solution of (\ref{eq:KKT_proj_row1_v2})-(\ref{eq:KKT_proj_row3_v2}).
It must be $\mu<0$; otherwise, since $\mathbf{X}-\mathbf{X}_{0}\succ\mathbf{0}$
{[}see (\ref{eq:X_0})], (\ref{eq:KKT_proj_row1_v2}) would lead to
$\mathbf{X}=\mathbf{0},$ which contradicts the power constraint in
(\ref{eq:KKT_proj_row3_v2}). Setting $\lambda=-\frac{1}{\mu}$, it
is easy to check that $(\mathbf{X},\lambda)$ satisfies (\ref{eq:KKT_rate_max_row1_v4})-(\ref{eq:KKT_rate_max_row3_v4}).

The system (\ref{eq:KKT_proj_row1_v2})-(\ref{eq:KKT_proj_row3_v2})
represents the KKT optimality conditions of problem (\ref{convex-projection})
with $\mathbf{X}_{0}$ defined in (\ref{eq:X_0}); which completes
the proof.\end{proof}\smallskip{}

Denoting by $\left[\mathbf{X}_{0}\right]_{\mathscr{Q}_{q}}$ the matrix
projection of $\mathbf{X}_{0}$ with respect to the Frobenius norm
onto the set $\mathscr{Q}_{q}$ defined in (\ref{set_Q_q})$-$the
solution to problem (P2) in (\ref{convex-projection}) with $P_{T}=P_{q}-$and
using Proposition \ref{Prop_MIMO_WF_eq} we have directly the following.%
\footnote{A more general expression of the waterfilling projection valid for
the general case of singular (possibly) rectangular channel matrices
is given in \cite{Scutari-GTMIMO}.%
}\smallskip{}

\begin{lemma} \label{Corollary-Multiuser_WF_Projection_MIMO}\emph{The
waterfilling operator} ${\mathsf{\mathbf{WF}}}_{q}\left(\mathbf{Q}_{-q}\right)$
\emph{in} (\ref{WF_MIMO_op}) \emph{can be equivalently written as}\begin{equation}
\mathsf{\mathbf{WF}}_{q}\left(\mathbf{Q}_{-q}\right)=\left[-\left(\mathbf{H}_{qq}^{H}\mathbf{R}_{-q}^{-1}(\mathbf{Q}_{-q})\mathbf{H}_{qq}\right)^{-1}\right]_{\mathscr{Q}_{q}},\label{WF_multiuser_projection}\end{equation}
\emph{where} $\mathscr{Q}_{q}$ \emph{is defined in} (\ref{set_Q_q}).\hfill $\square$\end{lemma}\smallskip{}

Comparing (\ref{Best_response_WF}) with (\ref{WF_multiuser_projection}),
it is straightforward to see that all the Nash equilibria of game
${\mathscr{G}}$ can be alternatively obtained as the fixed-points
of the mapping defined in (\ref{WF_multiuser_projection}):\begin{equation}
\mathbf{Q}_{q}^{\star}=\left[-\left(\mathbf{H}_{qq}^{H}\mathbf{R}_{-q}^{-1}(\mathbf{Q}_{-q}^{\star})\mathbf{H}_{qq}\right)^{-1}\right]_{\mathscr{Q}_{q}},\quad\forall q\in\Omega.\label{fixed-point_NE}\end{equation}

\noindent 

\bigskip{}

\noindent\addtocounter{rem}{1}\emph{Remark \therem}\textbf{\emph{
- }}\emph{Nonexpansive property of the MIMO waterfilling operator}\textbf{\emph{.}}
Thanks to the interpretation of MIMO waterfilling in (\ref{WF_MIMO_op})
as a projector, one can obtain the following nonexpansive property
of the waterfilling operator that will be used in the next section
to derive the contraction properties of the MIMO waterfilling mapping.
\smallskip{}

\noindent \begin{lemma} \label{NonExpansive-Lemma}\emph{Given} $q\in\Omega$,
let $\left[\cdot\right]_{\mathscr{Q}_{q}}$ \emph{denote the matrix
projection onto the convex set} $\mathscr{Q}_{q}$ \emph{with respect
to the Frobenius norm, as defined in} (\ref{convex-projection}).
\emph{Then}, \ $\left[\cdot\right]_{\mathscr{Q}_{q}}$\ \emph{satisfies
the following nonexpansive property:}\begin{equation}
\left\Vert \left[\mathbf{X}\right]_{\mathscr{Q}_{q}}-\left[\mathbf{Y}\right]_{\mathscr{Q}_{q}}\right\Vert _{F}\leq\left\Vert \mathbf{X}-\mathbf{Y}\right\Vert _{F},\;\forall\,\,\mathbf{X,Y\in\mathbb{C}}^{n_{T_{q}}\times n_{T_{q}}}.\label{NonExpansive_ineq}\end{equation}
 \hfill $\square$\end{lemma}\medskip{}

\subsubsection{Contraction property of MIMO multiuser waterfilling\label{Sec:Contraction-MIMO}}

Building on Lemmas \ref{Corollary-Multiuser_WF_Projection_MIMO} and
\ref{NonExpansive-Lemma}, we derive now sufficient conditions for
the waterfilling mapping to be a contraction, under a proper norm.
Our result is the natural extension of Theorem \ref{Theorem_contraction_SISO}
to the MIMO case.

As in the SISO case, we define first an appropriate block-maximum
norm for the multiuser waterfilling mapping. Given \begin{equation}
\mathbf{WF}(\mathbf{Q})=(\mathbf{WF}_{q}(\mathbf{Q}_{-q}))_{q\in\Omega}:{\mathscr{Q}\mapsto\mathscr{Q}},\label{MIMO_WF_mapping}\end{equation}
where ${\mathscr{Q}=\mathscr{Q}}_{1}\times\cdots\times{\mathscr{Q}}_{Q}$,
with ${\mathscr{Q}}_{q}$ and $\mathbf{WF}_{q}(\mathbf{Q}_{-q})$
defined in (\ref{set_Q_q}) and (\ref{WF_multiuser_projection}),
respectively, \ we introduce the following block-maximum norm on
$\mathbb{C}^{n\times n},$ with $n=n_{T_{1}}+\ldots+n_{T_{Q}},$ defined
as \cite{Bertsekas Book-Parallel-Comp} \begin{equation}
\left\Vert \mathsf{\mathbf{WF}}(\mathbf{Q})\right\Vert _{F,\text{block}}^{\mathbf{w}}\triangleq\max_{q\in\Omega}\frac{\left\Vert \mathbf{WF}_{q}(\mathbf{Q}_{-q})\right\Vert _{F}}{w_{q}},\text{ }\quad\label{block_max_weight_norm}\end{equation}
where $\left\Vert \mathbf{\cdot}\right\Vert _{F}$ is the Frobenius
norm and $\mathbf{w}\triangleq[w_{1},\ldots,w_{Q}]^{T}>\mathbf{0}$
is any positive weight vector. Finally, let $\mathbf{S}\in\mathbb{R}_{+}^{Q\times Q}$
be the nonnegative matrix defined as \begin{equation}
\left[\mathbf{S}\right]_{qr}\triangleq\left\{ \begin{array}{l}
\rho\left(\mathbf{H}_{rq}^{H}\mathbf{H}_{qq}^{-H}\mathbf{H}_{qq}^{-1}\mathbf{H}_{rq}\right),\\
0,\end{array}\begin{array}{l}
\text{if }r\neq q,\\
\text{otherwise.}\end{array}\right.\label{H_matrix}\end{equation}
where $\rho\left(\mathbf{A}\right)$ denotes the spectral radius%
\footnote{The spectral radius $\rho\left(\mathbf{A}\right)$ of the matrix $\mathbf{A}$
is defined as $\rho\left(\mathbf{A}\right)\triangleq\max\{|\lambda|:\lambda\in\sigma(\mathbf{A})\},$
with $\sigma(\mathbf{A})$ denoting the spectrum of $\mathbf{A}$
\cite{Horn85}.%
} of $\mathbf{A}$. The contraction property of the waterfilling mapping
is given in the following theorem. \smallskip{}

\begin{theorem}[Contraction property of mapping $\mathbf{WF}$] \label{Theorem_contraction}\emph{Given}
$\mathbf{w}\triangleq[w_{1},\ldots,w_{Q}]^{T}\mathbf{>0}$, \emph{the
mapping} $\mathbf{WF}$ \emph{defined in} (\ref{MIMO_WF_mapping})
\emph{is} Lipschitz continuous\emph{ on $\mathscr{Q}$}:\emph{ }\begin{equation}
\begin{array}{lc}
\hspace{-0.2cm}\left\Vert \mathsf{\mathbf{WF}}(\mathbf{Q}^{(1)})-\mathbf{WF}(\mathbf{Q}^{(2)})\right\Vert _{F,\text{block}}^{\mathbf{w}}\leq\hspace{-0.2cm} & \Vert\mathbf{S}\Vert_{\infty,\text{mat}}^{\mathbf{w}}\quad\quad\quad\quad\quad\quad\:\:\:\:\\
 & \hspace{-0.2cm}\times\left\Vert \mathbf{Q}^{(1)}-\mathbf{Q}^{(2)}\right\Vert _{F,\text{block}}^{\mathbf{w}},\quad\quad\end{array}\label{pseudo-contraction_def}\end{equation}
$\forall\mathbf{Q}^{(1)},\mathbf{Q}^{(2)}\in{\mathscr{Q}}$, \emph{where}
$\left\Vert \cdot\right\Vert _{F,\text{\emph{block}}}^{\mathbf{w}},$
$\left\Vert \cdot\right\Vert _{\infty,\text{mat}}^{\mathbf{w}}$ \emph{and}
$\mathbf{S}$ \emph{are defined in} (\ref{block_max_weight_norm}),
(\ref{H_max_weight_norm}), \emph{and} (\ref{H_matrix}), \emph{respectively},
\emph{and} $\mathscr{Q}\triangleq\mathscr{Q}_{1}\times\cdots\times\mathscr{Q}_{Q}$,
\emph{with} $\mathscr{Q}_{q}$ \emph{given in} (\ref{set_Q_q}). \emph{Furthermore},
\emph{if} \begin{equation}
\Vert\mathbf{S}\Vert_{\infty,\text{mat}}^{\mathbf{w}}<1\mathbf{,}\label{b_modulus_contraction}\end{equation}
\emph{then mapping} $\mathbf{WF}$ \emph{is a }\textit{\emph{block-contraction
with modulus}} $\alpha=\Vert\mathbf{S}\Vert_{\infty,\text{mat}}^{\mathbf{w}}.$
\end{theorem}\smallskip{}

\begin{proof} The proof of the theorem in the general case of arbitrary
channel matrices is quite involved \cite{Scutari-GTMIMO}. Here, we
focus only on the simpler case in which the direct channel matrices
$\{\mathbf{H}_{qq}\}_{q\in\Omega}$ are square and nonsingular. Under
this assumption, according to Lemma \ref{Corollary-Multiuser_WF_Projection_MIMO},
each component $\mathbf{WF}(\mathbf{Q}_{-q})$ of the mapping $\mathbf{WF}$
can be rewritten as in (\ref{WF_multiuser_projection}).

The proof consists in showing that the mapping $\mathbf{WF}$ satisfies
(\ref{pseudo-contraction_def}), with $\alpha=\Vert\mathbf{S}\Vert_{\infty,\text{mat}}^{\mathbf{w}}.$

Given $\mathbf{Q}^{(1)}=\left(\mathbf{Q}_{q}^{(1)},\ldots,\mathbf{Q}_{Q}^{(1)}\right)\in\mathscr{Q}$
and $\mathbf{Q}^{(2)}=\left(\mathbf{Q}_{1}^{(2)},\ldots,\mathbf{Q}_{Q}^{(2)}\right)\in\mathscr{Q},$
let define, for each $q\in\Omega$,\begin{eqnarray}
e_{\mathsf{WF}_{q}} & \triangleq & \left\Vert \mathsf{\mathbf{WF}}_{q}\left(\mathbf{Q}_{-q}^{(1)}\right)-\mathbf{WF}_{q}\left(\mathbf{Q}_{-q}^{(2)}\right)\right\Vert _{F},\label{e_q_and_e_T_q-1}\\
e_{q} & \triangleq & \left\Vert \mathbf{Q}_{q}^{(1)}-\mathbf{Q}_{q}^{(2)}\right\Vert _{F}.\label{e_q_and_e_T_q-2}\end{eqnarray}
Then, we have:\begin{align}
e_{\mathsf{WF}_{q}} & =\left\Vert \left[-\left(\mathbf{H}_{qq}^{H}\mathbf{R}_{q}^{-1}(\mathbf{Q}_{-q}^{(1)})\mathbf{H}_{qq}\right)^{-1}\right]_{{\mathscr{Q}q}}\right.\nonumber \\
 & \quad-\left.\left[-\left(\mathbf{H}_{qq}^{H}\mathbf{R}_{q}^{-1}(\mathbf{Q}_{-q}^{(2)})\mathbf{H}_{qq}\right)^{-1}\right]_{{\mathscr{Q}}_{q}}\right\Vert _{F}\medskip\label{Ineq_2}\\
 & \leq\left\Vert \left(\mathbf{H}_{qq}^{H}\mathbf{R}_{q}^{-1}(\mathbf{Q}_{-q}^{(1)})\mathbf{H}_{qq}\right)^{-1}-\left(\mathbf{H}_{qq}^{H}\mathbf{R}_{q}^{-1}(\mathbf{Q}_{-q}^{(2)})\mathbf{H}_{qq}\right)^{-1}\right\Vert _{F}\medskip\label{Ineq_3}\\
 & =\left\Vert \mathbf{H}_{qq}^{-1}\left(\sum\limits _{r\neq q\hfill}\mathbf{H}_{rq}\left(\mathbf{Q}_{r}^{(1)}-\mathbf{Q}_{r}^{(2)}\right)\mathbf{H}_{rq}^{H}\right)\mathbf{H}_{qq}^{-H}\right\Vert _{F}\medskip\label{ineq_6}\\
 & \leq\sum\limits _{r\neq q\hfill}\rho\left(\mathbf{H}_{rq}^{H}\mathbf{H}_{qq}^{-H}\mathbf{H}_{qq}^{-1}\mathbf{H}_{rq}\right)\left\Vert \left(\mathbf{Q}_{r}^{(1)}-\mathbf{Q}_{r}^{(2)}\right)\right\Vert _{F}\medskip\label{ineq_9}\\
 & \triangleq\sum\limits _{r\neq q\hfill\hfill}\left[\mathbf{S}\right]_{qr}\left\Vert \left(\mathbf{Q}_{r}^{(1)}-\mathbf{Q}_{r}^{(2)}\right)\right\Vert _{F}=\sum\limits _{r\neq q\hfill}\left[\mathbf{S}\right]_{qr}e_{r},\label{ineq_10}\end{align}
$\forall\mathbf{Q}^{(1)},\text{ }\mathbf{Q}^{(2)}\in{\mathscr{Q}}$
and $\forall q\in\Omega,$ where: (\ref{Ineq_2}) follows from (\ref{WF_multiuser_projection})
(Lemma \ref{Corollary-Multiuser_WF_Projection_MIMO}); (\ref{Ineq_3})
follows from the nonexpansive property of the projector in the Frobenius
norm as given in (\ref{NonExpansive_ineq}) (Lemma \ref{NonExpansive-Lemma});
(\ref{ineq_6}) follows from the nonsingularity of the channel matrices
$\{\mathbf{H}_{qq}\};$ (\ref{ineq_9}) follows from the triangle
inequality \cite{Horn85} and from \cite{Scutari-GTMIMO} \begin{equation}
\left\Vert \mathbf{A\mathbf{X}A}^{H}\right\Vert _{F}\leq\lambda_{\max}\left(\mathbf{A}^{H}\mathbf{A}\right)\left\Vert \mathbf{X}\right\Vert _{F},\label{inequality_lambda_max}\end{equation}
where $\mathbf{X=X}^{H}$ and $\mathbf{A\in\mathbb{C}}^{n\times m};$\ and
(\ref{ineq_10}) follows from the definition of matrix $\mathbf{S}$
in (\ref{H_matrix}).

Introducing the vectors \begin{equation}
\mathbf{e}_{\mathsf{WF}}\triangleq[e_{\mathsf{WF}_{1}},\ldots,e_{\mathsf{WF}_{Q}}]^{T},\mathbf{\quad}\text{and}\mathbf{\quad e}\triangleq[e_{1},\ldots,e_{Q}]^{T},\end{equation}
with $e_{\mathsf{WF}_{q}}$ and$\ e_{q}$ defined in (\ref{e_q_and_e_T_q-1})
and (\ref{e_q_and_e_T_q-2}), respectively, the set of inequalities
in (\ref{ineq_10}) can be rewritten in vector form as\begin{equation}
\mathbf{0}\leq\mathbf{e}_{\mathsf{WF}}\leq\mathbf{Se},\quad\forall\mathbf{Q}^{(1)},\text{ }\mathbf{Q}^{(2)}\in{\mathscr{Q}}\text{.}\label{vector_e_}\end{equation}
Using the weighted maximum norm $\left\Vert \mathbf{\cdot}\right\Vert _{\infty,\text{vec}}^{\mathbf{w}}$
defined in (\ref{weighted_infinity_vector_norm}) in combination with
(\ref{vector_e_}), we have\begin{equation}
\left\Vert \mathbf{e}_{\mathsf{WF}}\right\Vert _{\infty,\text{vec}}^{\mathbf{w}}\leq\left\Vert \mathbf{Se}\right\Vert _{\infty,\text{vec}}^{\mathbf{w}}\leq\left\Vert \mathbf{S}\right\Vert _{\infty,\text{mat}}^{\mathbf{w}}\left\Vert \mathbf{e}\right\Vert _{\infty,\text{vec}}^{\mathbf{w}},\label{e_t_contraction}\end{equation}
$\forall\mathbf{Q}^{(1)},\mathbf{Q}^{(2)}\in{\mathscr{Q}}$ and $\mathbf{w}>\mathbf{0}$,
where $\left\Vert \cdot\right\Vert _{\infty,\text{mat}}^{\mathbf{w}}$
is the matrix norm induced by the vector norm $\left\Vert \mathbf{\cdot}\right\Vert _{\infty,\text{vec}}^{\mathbf{w}}$
in (\ref{weighted_infinity_vector_norm}) and defined in (\ref{H_max_weight_norm})
\cite{Horn85}. Finally, using (\ref{e_t_contraction}) and (\ref{block_max_weight_norm}),
we obtain,\begin{equation}
\begin{array}{l}
\left\Vert \mathsf{\mathbf{WF}}\mathbf{(\mathbf{Q}}^{(1)}\mathbf{)}-\mathbf{WF}\mathbf{(\mathbf{Q}}^{(2)}\mathbf{)}\right\Vert _{F,\text{block}}^{\mathbf{w}}=\left\Vert \mathbf{e}_{\mathsf{WF}}\right\Vert _{\infty,\text{vec}}^{\mathbf{w}}\qquad\qquad\qquad\qquad\medskip\\
\qquad\qquad\qquad\quad\qquad\qquad\leq\left\Vert \mathbf{S}\right\Vert _{\infty,\text{mat}}^{\mathbf{w}}\left\Vert \mathbf{\mathbf{Q}}^{(1)}-\mathbf{\mathbf{Q}}^{(2)}\right\Vert _{F,\text{block}}^{\mathbf{w}},\end{array}\label{eq:eq-final-proof}\end{equation}
$\forall\mathbf{Q}^{(1)},\mathbf{Q}^{(2)}\in{\mathscr{Q}}$ and $\forall\mathbf{w}>\mathbf{0},$
which leads to a block-contraction for the mapping $\mathbf{WF}$
if $\left\Vert \mathbf{S}\right\Vert _{\infty,\text{mat}}^{\mathbf{w}}<1,$
implying condition (\ref{b_modulus_contraction}). \end{proof}

\section{Existence and Uniqueness of the NE\label{Existence-Uniqueness}}

Using results obtained in the previous section, we can now study game
${\mathscr{G}}\ $ and derive conditions for existence and uniqueness
of the NE$,$ as given next.\smallskip{}

\begin{theorem} \label{th:existence_uniqueness_NE_MIMO}\emph{Game}
${\mathscr{G}}\ $ \emph{always admits a NE, for any set of channel
matrices and transmit power of the users. Furthermore, the NE is unique
if}\vspace{-0.3cm}
\begin{equation}
\rho\left(\mathbf{S}\right)<1,\tag{C1}\label{SF_MIMO}\end{equation}
\emph{where} $\mathbf{S}$ \emph{is defined in }(\ref{H_matrix}).
\end{theorem}\smallskip{}

\begin{proof} According to the interpretation of the waterfilling
mapping $\mathbf{WF}$ in (\ref{WF_MIMO_op}) as a projector (cf.
Lemma \ref{Corollary-Multiuser_WF_Projection_MIMO}), the existence
of a NE of game ${\mathscr{G}}$ is guaranteed by the existence of
a solution of the fixed-point equation (\ref{fixed-point_NE}). Invoking
Theorem \ref{Theo_existemce-uniq}(a), the existence of a fixed-point
follows from the continuity of the waterfilling projector (\ref{WF_multiuser_projection})
on $\mathscr{Q}$, for any given set of channel matrices $\{\mathbf{H}_{rq}\}_{r,q\in\Omega}$
(implied from the continuity of the projection operator \cite[Proposition 3.2c]{Bertsekas Book-Parallel-Comp}
and the continuity of each $\mathbf{R}_{-q}^{-1}(\mathbf{Q}_{-q})$
on $\mathscr{Q}_{-q}$%
\footnote{This result can be proved using \cite[Theorem 10.7.1]{Campbell-Meyer-book}.%
}), and from the convexity and compactness of the joint admissible
strategy set $\mathscr{Q}.$%
\footnote{According to \cite[Theorem 1]{Rosen}, the existence of a NE of game
${\mathscr{G}}$ can also be proved showing that the game is a concave
game: 1) The set $\mathscr{Q}_{q}$ of feasible strategy profiles
of each player $q$ is compact and convex; and 2) The payoff function
of each player $q$ is continuous in $\mathbf{Q}\in\mathscr{Q}$ and
concave in $\mathbf{Q}_{q}\in\mathscr{Q}_{q}$, for any given $\mathbf{Q}_{-q}\in\mathscr{Q}_{-q}$
(this follows from the concavity of the log function \cite{Boyd-book}). %
}

According to Theorem \ref{Theo_existemce-uniq}(b), a sufficient condition
for the uniqueness of the NE of game ${\mathscr{G}}$ is that the
waterfilling mapping $\mathbf{WF}$ in (\ref{WF_MIMO_op}) be a contraction
with respect to some norm. It follows from Theorem \ref{Theorem_contraction}
that $\mathbf{WF}$ is a block-contraction if condition (\ref{b_modulus_contraction})
is satisfied for some $\mathbf{w}>\mathbf{0}$. Since $\mathbf{S}$
in (\ref{b_modulus_contraction}) is a nonnegative matrix, there exists
a positive vector $\overline{\mathbf{w}}$ such that \cite[Corollary 6.1]{Bertsekas Book-Parallel-Comp}\begin{equation}
\left\Vert \mathbf{S}\right\Vert _{\infty,\limfunc{mat}}^{\overline{\mathbf{w}}}<1\quad\Leftrightarrow\quad\rho\left(\mathbf{S}\right)<1,\label{eq_nor_spectral_radius}\end{equation}
 which proves the sufficiency of (\ref{SF_MIMO}). \end{proof}

\noindent \smallskip{}

To give additional insight into the physical interpretation of sufficient
conditions for the uniqueness of the NE, we provide the following
corollary of Theorem \ref{th:existence_uniqueness_NE_MIMO}.

\begin{corollary} \label{Corollary-C1-C2_MIMO}\emph{A sufficient
condition for} (\ref{SF_MIMO}) \emph{is given by one of the two following
set of conditions}:\begin{equation}
\dfrac{1}{w_{q}}\text{ }\!\!\dsum\limits _{r\neq q}\rho\left(\mathbf{H}_{rq}^{H}\mathbf{H}_{qq}^{-H}\mathbf{H}_{qq}^{H}\mathbf{H}_{rq}\right)w_{r}<1,\text{ }\forall q\in\Omega,\hspace{-0.15cm}\vspace{-0.2cm}\medskip\tag{C2}\label{SF_for_C1_a}\end{equation}
\begin{equation}
\dfrac{1}{w_{r}}\!\!\dsum\limits _{q\neq r}\rho\left(\mathbf{H}_{rq}^{H}\mathbf{H}_{qq}^{-H}\mathbf{H}_{qq}^{H}\mathbf{H}_{rq}\right)w_{q}<1,\text{ }\forall r\in\Omega,\tag{C3}\label{SF_for_C1_b}\end{equation}
\emph{where} $\mathbf{w}\triangleq\lbrack w_{1},\ldots,w_{Q}]^{T}$
\emph{is a positive vector.}\hfill $\square$ \end{corollary}\addtocounter{rem}{1}\smallskip{}

\noindent\emph{Remark {\therem} - Physical interpretation of uniqueness
conditions.}\textbf{ }Looking at conditions (\ref{SF_for_C1_a})-(\ref{SF_for_C1_b}),
it turns out, as expected, that the uniqueness of a NE is ensured
if the interference among the links is sufficiently small. The importance
of conditions (\ref{SF_for_C1_a})-(\ref{SF_for_C1_b}) is that they
quantify how small the interference must be to guarantee that the
equilibrium is indeed unique. Specifically, condition (\ref{SF_for_C1_a})
can be interpreted as a constraint on the maximum amount of interference
that each receiver can tolerate, whereas (\ref{SF_for_C1_b}) introduces
an upper bound on the maximum level of interference that each transmitter
is allowed to generate. This result agrees with the intuition that,
as the MUI becomes negligible, the rates of the users become decoupled
and then the rate-maximization problem in (\ref{Rate-matrix-game})
 for each user admits a unique solution.\smallskip{}

\noindent \addtocounter{rem}{1}\emph{Remark {\therem} - Special
cases.}\textbf{ }Conditions in Theorem \ref{th:existence_uniqueness_NE_MIMO}
and Corollary \ref{Corollary-C1-C2_MIMO} for the uniqueness of the
NE can be applied to \emph{arbitrary} MIMO interference systems,%
\footnote{Recall that here we have concentrated on square nonsingular (direct)
MIMO channels. However, condition (\ref{SF_MIMO}) can be generalized
to the case of rectangular channel matrices \cite{Scutari-GTMIMO}.%
} irrespective of the specific structure of channel matrices. Interestingly,
most of the conditions known in the literature \cite{Yu}-\cite{Huang-Cendrillon}
for the rate-maximization game in \emph{SISO frequency-selective}
interference channels and OFDM transmission come naturally from (\ref{SF_MIMO})
as special cases. In fact, using the Toeplitz and circulant structure
of the channel matrices $\mathbf{H}_{rq}=\mathbf{WD}_{rq}\mathbf{W}^{H}$
(cf. Section \ref{Sec:Frequency-Selective-case}), matrix $\mathbf{S}$
in the uniqueness condition (\ref{SF_MIMO}), defined in (\ref{H_matrix}),
reduces to matrix $\mathbf{S}^{\max}$ defined in (\ref{Hmax}), showing
that our uniqueness condition coincides with those given in \cite{Scutari-Part II,Scutari-IT-08}
and enlarges those obtained in \cite{Yu}-\cite{Scutari-Barbarossa-SPAWC03}
and \cite{Tse,Huang-Cendrillon}. Observe that condition (\ref{SF_MIMO}),
with $\mathbf{S}^{\max}$ defined in (\ref{Hmax}), can be further
weakened by computing the {}``$\max$\textquotedblright\ over a
subset of $\{1,\ldots,N\},$ obtained from $\{1,\ldots,N\}$ by removing
the subcarrier indexes where each user will never transmit, for any
set of channel realizations and interference profile \cite{Scutari-Part I,Scutari-Part II}.
An algorithm to compute such a set is given in \cite{Scutari-Part I}.

\noindent \indent Recently, in \cite{Arslan_etal}, the authors studied
the game $\mathscr{G}$ and proved using \cite[Theorem 2.2]{Rosen}
that the NE of the game is unique if the MUI at each receiver $q$,
measured by the interference-to-noise ratios $\{P_{r}/\sigma_{q}^{2}\}_{r\neq q}$
where $\sigma_{q}^{2}$ is the variance of the thermal noise at receiver
$q$, is smaller than a given \emph{unspecified} threshold. Differently
from \cite{Arslan_etal}, our results provide a set of sufficient
conditions that can be checked in practice, since they explicitly
quantify how strong the MUI must be to guarantee the uniqueness of
the NE.

\section{MIMO Asynchronous Iterative Waterfilling Algorithm\label{Section_Distributed Algorithms}}

\noindent According to the framework developed in Section \ref{Sec:Unified_Tool},
to reach the Nash equilibria of game ${\mathscr{G}}$, one can use
an instance of the totally asynchronous scheme of \cite{Bertsekas Book-Parallel-Comp}
(cf. Section \ref{Sec:fixed_point_convergence}), based on the waterfilling
mapping (\ref{WF_MIMO_op}), called asynchronous Iterative WaterFilling
Algorithm (IWFA) \cite{Scutari-GTMIMO}. In the asynchronous IWFA,
all the users maximize their own rate in a \emph{totally asynchronous}
way via the single user waterfilling solution (\ref{WF_MIMO_op}).
According to this asynchronous procedure, some users are allowed to
update their strategy more frequently than the others, and they might
perform these updates using \emph{outdated}  information on the interference
caused by the others. We show in the following that, whatever the
asynchronous mechanism is, such a procedure converges to a stable
NE of the game, under the same sufficient conditions guaranteeing
the uniqueness of the equilibrium given in Theorem \ref{th:existence_uniqueness_NE_MIMO}.

To provide a formal description of the proposed asynchronous IWFA,
we need the following preliminary definitions. We assume, w.l.o.g.,
that the set of times at which one or more users update their strategies
is the discrete set $\mathcal{T}=\mathbb{N}_{+}=\left\{ 0,1,2,\ldots\right\} .$
Let $\mathbf{Q}_{q}^{(n)}$ denote the covariance matrix of the vector
signal transmitted by user $q$ at the $n$-th iteration, and let
$\mathcal{T}_{q}\subseteq\mathcal{T}$ \ denote the set of times
$n$ at which $\mathbf{Q}_{q}^{(n)}$ is updated (thus, at time $n\notin\mathcal{T}_{q},$
$\mathbf{Q}_{q}^{(n)}$ is left unchanged). Let $\tau_{r}^{q}(n)$
denote the most recent time at which the interference from user $r$
is perceived by user $q$ at the $n$-th iteration (observe that $\tau_{r}^{q}(n)$
satisfies $0\leq\tau_{r}^{q}(n)\leq n$). Hence, if user $q$ updates
his own covariance matrix at the $n$-th iteration, then he chooses
his optimal $\mathbf{Q}_{q}^{(n)}$, according to (\ref{WF_MIMO_op}),
and using the interference level caused by \begin{eqnarray}
\mathbf{Q}_{-q}^{(\mathbf{\tau}^{q}(n))} & \triangleq & \left(\mathbf{Q}_{1}^{(\tau_{1}^{q}(n))},\ldots,\mathbf{Q}_{q-1}^{(\tau_{q-1}^{q}(n))},\right.\nonumber \\
 &  & \left.\;\;\mathbf{Q}_{q+1}^{(\tau_{q+1}^{q}(n))},\ldots,\mathbf{Q}_{Q}^{(\tau_{Q}^{q}(n))}\right).\label{p_q_interference}\end{eqnarray}

The overall system is said to be totally asynchronous if the following
weak assumptions are satisfied for each $q$ \cite{Bertsekas Book-Parallel-Comp}:
A1) $0\leq\tau_{r}^{q}(n)\leq n;$ A2) $\lim_{k\rightarrow\infty}\tau_{r}^{q}(n_{k})=+\infty;$
and A3) $\left\vert \mathcal{T}_{q}\right\vert =\infty;$ where $\{n_{k}\}$
is a sequence of elements in $T_{q}$ that tends to infinity. Assumption
(A1)-(A3) are standard in asynchronous convergence theory \cite{Bertsekas Book-Parallel-Comp},
and they are fulfilled in any practical implementation. In fact, (A1)
simply indicates that, in the current iteration $n$, each user $q$
can use only interference vectors $\mathbf{Q}_{-q}^{(\mathbf{\tau}^{q}(n))}$
allocated by others in previous iterations (to preserve causality).
Assumption (A2) states that, for any given iteration index $n_{1},$
values of the components of $\mathbf{Q}_{-q}^{(\mathbf{\tau}^{q}(n))}$
in (\ref{p_q_interference}) generated prior to $n_{1},$ will not
be used in the updates of $\mathbf{Q}_{q}^{(n)}$ after a sufficiently
long time $n_{2};$ this guarantees that old information is eventually
purged from the system. Finally, assumption (A3) indicates that no
user fails to update his own strategy as time $n$ goes on.

Using the above notation, the asynchronous IWFA is formally described
in Algorithm 1.\medskip{}

\begin{algo}{MIMO Asynchronous IWFA} SSet $n=0$ and $\mathbf{Q}_{q}^{(0)}=$
any feasible covariance matrix; \\
\texttt{for} $n=0:\mathrm{N_{it}}$ \\
 \begin{equation}
\,\,\,\,\mathbf{Q}_{q}^{(n+1)}=\left\{ \begin{array}{ll}
\mathbf{WF}_{q}\left(\mathbf{Q}_{-q}^{(\mathbf{\tau}^{q}(n))}\right), & \text{if }n\in\mathcal{T}_{q},\\
\mathbf{Q}_{q}^{(n)}, & \text{otherwise};\end{array}\right.\quad\forall q\in\Omega\label{AIWFA}\end{equation}
 \\
 \texttt{end} \end{algo}

\bigskip{}

It follows directly from Theorems \ref{unified-set-of-conditions}
and \ref{Theorem_contraction} that convergence of the algorithm is
guaranteed under the following sufficient conditions.\smallskip{}

\begin{theorem} \label{Theo-AIWFA_MIMO} \emph{Suppose that condition}
\ (\ref{SF_MIMO}) \emph{in Theorem} \ref{th:existence_uniqueness_NE_MIMO}
\emph{is satisfied. Then, as} $\mathrm{N_{it}}$ $\rightarrow\infty,$
\emph{the asynchronous IWFA, described in Algorithm 1, converges to
the unique NE of game} ${\mathscr{G}}$ \emph{for any set of feasible
initial conditions and updating schedule.} \hfill $\square$\end{theorem}\smallskip{}

\noindent \addtocounter{rem}{1}\emph{Remark {\therem\ - Global
convergence and robustness of the algorithm}}\textbf{\emph{.}} Even
though the rate maximization game ${\mathscr{G}}$ and the consequent
waterfilling mapping (\ref{WF_MIMO_op}) are nonlinear, condition
(\ref{SF_MIMO}) guarantees the \emph{global} convergence of the asynchronous
IWFA. Observe that Algorithm 1 contains as special cases a plethora
of algorithms, each one obtained by a possible choice of the scheduling
of the users in the updating procedure (i.e., the parameters $\{\tau_{r}^{q}(n)\}$
and $\{\mathcal{T}_{q}\}$). Two special cases are the \emph{sequential}
and the \emph{simultaneous} MIMO IWFA, where the users update their
own strategies \emph{sequentially} and \emph{simultaneously,} respectively.
The important result stated in Theorem \ref{Theo-AIWFA_MIMO} is that
all the algorithms resulting as special cases of the asynchronous
IWFA are guaranteed to reach the unique NE of the game, under the
same set of convergence conditions (provided that (A1)-(A3) are satisfied),
since conditions in (\ref{SF_MIMO}) do not depend on the particular
choice of $\{\mathcal{T}_{q}\}$ and $\{\tau_{r}^{q}(n)\}.$

\smallskip{}

\noindent\addtocounter{rem}{1}\emph{Remark {\therem\ - Distributed
nature of the algorithm.}\textbf{ }}Since the asynchronous IWFA is
based on the waterfilling solution (\ref{WF_MIMO_op}), it can be
implemented in a distributed way, where each user, to maximize his
own rate, only needs to measure the covariance matrix of the overall
interference-plus-noise and waterfill over this matrix. More interestingly,
according to the asynchronous scheme, the users may update their strategies
using a potentially outdated version of the interference and, furthermore,
some users are allowed to update their covariance matrix more often
than others, without affecting the convergence of the algorithm. These
features strongly relax the constraints on the synchronization of
the users' updates with respect to those imposed, for example, by
the simultaneous or sequential updating schemes.

\smallskip{}

\noindent\addtocounter{rem}{1}\emph{Remark {\therem\ - Well-known
cases.}\textbf{ }}The MIMO asynchronous IWFA, described in Algorithm
1 is the natural generalization of the asynchronous IWFA proposed
in \cite{Scutari-IT-08}, to solve the rate-maximization game in Gaussian
\emph{SISO frequency-selective} parallel interference channels\emph{.}
Algorithm in \cite{Scutari-IT-08} can be in fact obtained directly
from Algorithm 1 using the following equivalences: $\mathbf{Q}_{q}\Leftrightarrow\mathbf{p}_{q}$,
$\mathbf{WF}_{q}\left(\mathbf{\cdot}\right)\Leftrightarrow\mathsf{wf}_{q}\left(\mathbf{\cdot}\right),$
and $\text{ }\mathscr{Q}_{q}\Leftrightarrow{\mathscr{P}}_{q},$ where
$\mathbf{WF}_{q}\left(\mathbf{\cdot}\right),$ $\mathsf{wf}_{q}\left(\mathbf{\cdot}\right)$,
$\mathscr{Q}_{q},$ and ${\mathscr{P}}_{q}$ are defined in (\ref{WF_MIMO_op}),
(\ref{WF_OFDM}), (\ref{set_Q_q}), and (\ref{set_P_q}), respectively.
Similarly, the well-known sequential IWFA \cite{Yu}-\cite{Scutari_Thesis},
\cite{Scutari-Part II} and simultaneous IWFA \cite{Scutari_Thesis}-\cite{Sung-JSAC07},
\cite{Scutari-Part II} proposed in the literature are special cases
of Algorithm 1, using the above equivalences.

\section{Numerical Results}

\noindent \label{Num_Res} In this section, we first provide some
numerical results illustrating the benefits of MIMO transceivers in
the multiuser context. Then, we compare some of the proposed algorithms
in terms of convergence speed. \smallskip{}

\begin{figure}[tbph]
\hspace{-0.39cm}\includegraphics[height=5.5cm]{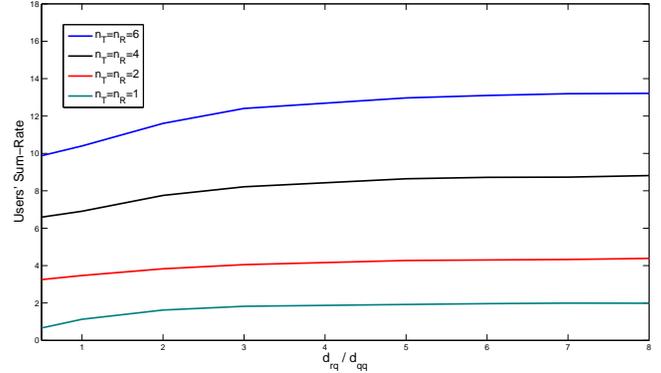} 
\par

\vspace{-0.4cm}

\caption{{\protect{\small Sum-Rate of the users versus $d_{rq}/d_{qq}$;
$d_{rq}=d_{qr},$ $d_{rr}=d_{qq}=1$, $r=1,2,$ $Q=2,$ $\gamma=2.5,$
$P_{1}/\sigma_{1}^{2}=P_{2}/\sigma_{2}^{2}=5$dB, $L_{h}=6,$ $N=16.$}}}

\label{Fig:MIMO_SISO} 
\end{figure}

\noindent \emph{Example 1}\textbf{\emph{ $-$ }}\emph{MIMO vs. SISO.}
MIMO systems have shown great potential for providing high spectral
efficiency in both isolated, single-user, wireless links without interference
or multiple access and broadcast channels. Here we quantifies, by
simulations, this potential gain for MIMO interference systems. In
Figure \ref{Fig:MIMO_SISO}, we plot the sum-rate of a two-user frequency-selective
MIMO system as a function of the inter-pair distance among the links,
for different number of transmit/receive antennas. The rate curves
are averaged over $500$ independent channel realizations, whose taps
are simulated as i.i.d. Gaussian random variables with zero mean and
unit variance. For the sake of simplicity, the system is assumed to
be symmetric, i.e., the transmitters have the same power budget and
the interference links are at the same distance (i.e., $d_{rq}=d_{qr},\,\,\forall q,r$),
so that the cross channel gains are comparable in average sense. The
path loss $\gamma$ is assumed to be $\gamma=2.5.$

From the figure one infer that, as for isolated single-user systems
or multiple access/broadcast channels, also in MIMO interference channels,
increasing the number of antennas at both the transmitter and the
receiver side leads to a better performance. The interesting result,
coming from Figure \ref{Fig:MIMO_SISO}, is that the incremental gain
due to the use of multiple transmit/receive antennas is almost independent
of the interference level in the system, since the MIMO (incremental)
gains in the high-interference case (small values of $d_{rq}/d_{qq}$)
almost coincide with the corresponding (incremental) gains obtained
in the low-interference case (large values of $d_{rq}/d_{qq}$), at
least for the system simulated in Figure \ref{Fig:MIMO_SISO}. This
desired property is due to the fact that the MIMO channel provides
more degrees of freedom for each user than those available in the
SISO channel, that can be explored to find out the best partition
of the available resources for each user, possibly cancelling the
MUI.%
\smallskip{}

\begin{figure}[tbph]
\begin{centering}
\includegraphics[height=6cm]{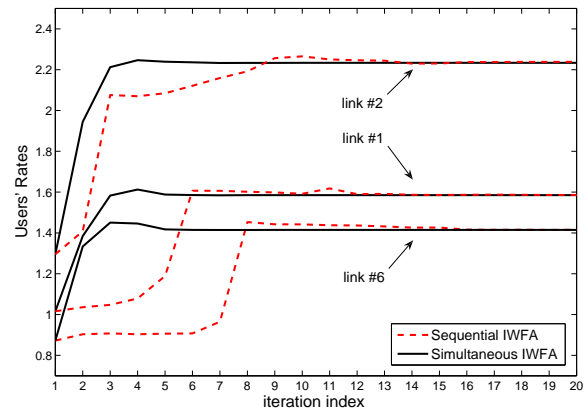} 
\par\end{centering}

\vspace{-0.3cm}

\caption{{\protect{\small Rates of the links versus iterations: sequential
IWFA (dashed line curves) and simultaneous IWFA (solid line curves);
$Q=8,$ $\gamma=2.5,$ $P_{q}/\sigma_{q}^{2}=5$dB, $\forall q\in\Omega$,
$L_{h}=6,$ $N=16.$}}}

\label{Fig_SIWFA-IWFA} 
\end{figure}

\noindent \emph{Example 2}\textbf{\emph{ $-$ }}\emph{Sequential vs.
simultaneous IWFA.} In Figure \ref{Fig_SIWFA-IWFA} we compare the
performance of the sequential and simultaneous IWFA, in terms of convergence
speed, for a given set of MIMO channel realizations. We consider a
cellular network composed by $7$ (regular) hexagonal cells, sharing
the same spectrum. Hence, simultaneous transmissions of different
cells can interfere with each other. The Base Stations (BS) and the
Mobile Terminals (MT) are equipped with $4$ antennas. For the sake
of simplicity, we assume that in each cell there is only one active
link, corresponding to the transmission from the BS (placed at the
center of the cell) to a MT placed in a corner of the cell. According
to this geometry, each MT receives an useful signal that is comparable
in average sense with the interference signal transmitted by the BSs
of two adjacent cells. The overall network is thus stitched out of
eight $4\times4$ MIMO interference wideband channels, according to
(\ref{vector I/O}).

In Figure \ref{Fig_SIWFA-IWFA}, we show the rate evolution of the
links of three cells corresponding to the sequential IWFA and simultaneous
IWFA as a function of the iteration index $n$ . To make the figure
not excessively overcrowded, we plot only the curves of $3$ out of
$8$ links. As expected, the sequential IWFA is slower than the simultaneous
IWFA, especially if the number of active links $Q$ is large, since
each user is forced to wait for all the users scheduled in advance,
before updating his own power allocation. The same qualitative behavior
has been observed changing the channel realizations and the number
of antennas.%

\section{Conclusions\label{Sec:Conclusion}}

\noindent In this paper we have considered the competitive maximization
of mutual information in noncooperative interfering networks in a
fully distributed fashion, based on game theory. We have provided
a unified view of main results obtained in the past seven years, showing
that the proposed approaches, even apparently different, can be unified
by our interpretation of the watefilling solution as a proper projection
onto a polyhedral set. Building on this interpretation, we have shown
how to apply standard results from fixed-point and contraction theory
to the rate maximization game in SISO frequency-selective channels,
in order to obtain a unified set of sufficient conditions guaranteeing
the uniqueness of the NE and the convergence of totally asynchronous
distributed algorithms. The proposed framework has also been generalized
to the (square) MIMO case. The obtained results are the natural generalization
of those obtained in the SISO case.


\begin{thebibliography}{10}
\bibitem{Starr-Cioffi Book}T. Starr, J. M. Cioffi, and P. J. Silverman,
\textit{Understanding Digital Subscriber Line Technology}, Prentice
Hall, NJ, 1999.

\bibitem{Goldsmith-Wicker} A. J. Goldsmith and S. B. Wicker, {}``Design
Challenges for Energy-Constrained Ad Hoc Wireless Networks,\textquotedblright
\textit{IEEE Wireless Communications Magazine}, vol. 9, no. 4, pp.
8-27, August 2002.

\bibitem{Akyildiz-Wang} I. F. Akyildiz and X. Wang, {}``A Survey
on Wireless Mesh Networks,\textquotedblright \textit{IEEE Communications
Magazine}, vol. 43, no. 9, pp. 23-30, September 2005.

\bibitem{Haykin} S. Haykin, {}``Cognitive Radio: Brain-Empowered
Wireless Communications,\textquotedblright \textit{IEEE Jour. on
Selected Areas in Communications}, vol. 23, no. 2, pp. 201-220, February
2005.

\bibitem{Osborne} M. J. Osborne and A. Rubinstein, \textit{A Course
in Game Theory}, MIT Press, 1994.

\bibitem{Aubin-book} J. P. Aubin, \textit{Mathematical Method for
Game and Economic Theory}, Elsevier, Amsterdam, 1980.

\bibitem{Yu} W. Yu, G. Ginis, and J. M. Cioffi, {}``Distributed
Multiuser Power Control for Digital Subscriber Lines,\textquotedblright{}
\textit{IEEE Jour. on Selected Areas in Communications}, vol. 20,
no. 5, pp. 1105-1115, June 2002.

\bibitem{ChungISIT03} S. T. Chung, S. J. Kim, J. Lee, and J. M. Cioffi,
{}``A Game-theoretic Approach to Power Allocation in Frequency-selective
Gaussian Interference Channels,\textquotedblright{} in \textit{Proc.
of the 2003 IEEE International Symposium on Information Theory (ISIT
2003)}, p. 316, June 2003.

\bibitem{Yamashitay-Luo} N. Yamashita and Z. Q. Luo, {}``A Nonlinear
Complementarity Approach to Multiuser Power Control for Digital Subscriber
Lines,\textquotedblright \textit{Optimization Methods and Software,}
vol. 19, no. 5, pp. 633--652, October 2004.

\bibitem{Scutari-Barbarossa-ICASSP} G. Scutari, S. Barbarossa, and
D. Ludovici, {}``On the Maximum Achievable Rates in Wireless Meshed
Networks: Centralized versus Decentralized solutions,\textquotedblright
in \textit{Proc. of the 2004 IEEE Int. Conf. on Acoustics, Speech,
and Signal Processing (ICASSP-2004)}, May 2004.

\bibitem{Scutari-Barbarossa-SPAWC03} G. Scutari, S. Barbarossa, and
D. Ludovici, {}``Cooperation Diversity in Multihop Wireless Networks
Using Opportunistic Driven Multiple Access,\textquotedblright{}
in \textit{Proc. of the 2003 IEEE Workshop on Sig. Proc. Advances
in Wireless Comm., (SPAWC-2003)}, pp. 170-174, June 2003.

\bibitem{Luo-Pang}Z.-Q. Luo and J.-S. Pang, {}``Analysis of Iterative
Waterfilling Algorithm for Multiuser Power Control in Digital Subscriber
Lines,\textquotedblright{} \textit{EURASIP Jour. on Applied Signal
Processing}, May 2006.

\bibitem{Tse} R. Etkin, A. Parekh, and D. Tse, {}``Spectrum Sharing
for Unlicensed Bands,\textquotedblright{} in \textit{Proc. of the
Allerton Conference on Commuication, Control, and Computing}, Monticello,
IL, September 28-30, 2005.

\bibitem{Scutari_Thesis} G. Scutari, \textit{Competition and Cooperation
in Wireless Communication Networks}, PhD. Dissertation, University
of Rome, {}``La Sapienza\textquotedblright,{} November 2004.

\bibitem{Huang-Cendrillon} R. Cendrillon, J. Huang, M. Chiang and
M. Moonen, {}``Autonomous Spectrum Balancing for Digital Subscriber
Lines,\textquotedblright{} \emph{IEEE Trans. on Signal Processing},
vol. 55, no. 8, p. 4241--4257, Aug. 2007.

\bibitem{Sung-JSAC07}K. W. Shum, K.-K. Leung, C. W. Sung, {}``Convergence
of Iterative Waterfilling Algorithm for Gaussian Interference Channels,\textquotedblright{}
\emph{IEEE Jour. on Selected Area in Communications}, vol. 25, no
6, pp. 1091-1100, Aug. 2007. 

\bibitem{Scutari-ICASSP07} G. Scutari, D. P. Palomar, and Sergio
Barbarossa, {}``Distributed Totally Asynchronous Iterative Waterfilling
for Wideband Interference Channel with Time/Frequency Offset,\textquotedblright{}
in Proc. of the \textit{IEEE Int. Conf. on Acoustics, Speech, and
Signal Processing (ICASSP)}, Honolulu, Hawaii, USA, April 15-20, 2007.

\bibitem{Scutari-Part I} G. Scutari, D. P. Palomar, and S. Barbarossa,
{}``Optimal Linear Precoding Strategies for Wideband Non-Cooperative
Systems based on Game Theory-Part I: Nash Equilibria,\textquotedblright{}
\textit{IEEE Trans. on Signal Processing,} Vol. 56, no. 3, pp. 1230-1249,
March 2008. 

\bibitem{Scutari-Part II} G. Scutari, D. P. Palomar, and S. Barbarossa,
{}``Optimal Linear Precoding Strategies for Wideband Non-Cooperative
Systems based on Game Theory-Part II: Algorithms,\textquotedblright{}
\textit{IEEE Trans. on Signal Processing,} , Vol. 56, no. 3, pp. 1250-1267,
March 2008. See also Proc. of . \textit{IEEE International Symposium
on Information Theory (ISIT)}, Seattle, WA, USA, July 9-14, 2006.

\bibitem{Scutari-IT-08}G. Scutari, D. P. Palomar, and S. Barbarossa,
{}``Asynchronous Iterative Waterfilling for Gaussian Frequency-Selective
Interference Channels,\textquotedblright{} to appear on \textit{IEEE
Trans. on Information Theory, }\textit{\emph{July 2008. See also }}Proc.
\textit{IEEE Workshop on Signal Proc. Advances in Wireless Commun.
(SPAWC 2006)}, Cannes, France, July 2-5, 2006\textit{\emph{; and Proc.
of }}\textit{IEEE} \textit{Information Theory and Applications (ITA)
Workshop}, San Diego, CA, USA, Jan. 29 - Feb. 2, 2007.

\bibitem{Scutari-GTMIMO} G. Scutari, D. P. Palomar, and S. Barbarossa,
{}``The  MIMO Iterative Waterfilling Algorithm,\textquotedblright{}
 submitted to \textit{IEEE Trans. on Signal
Processing}. See also Proc. of the \textit{IEEE Int. Conf. on Acoustics,
Speech, and Signal Processing (ICASSP)}, Las Vegas, USA, March 30
- April 4, 2008.

\bibitem{Bertsekas Book-Parallel-Comp}D. P Bertsekas and J.N. Tsitsiklis,
\textit{Parallel and Distributed Computation: Numerical Methods},
Athena Scientific, 2nd Ed., 1989.

\bibitem{Yates-Jsac} R. D. Yates, {}``A Framework for Uplink Power
Control in Cellular Radio Systems,\textquotedblright{} \textit{IEEE
Jour. on Selected Area in Communications}, vol. 13, no 7, pp. 1341-1347,
September 1995.

\bibitem{Ortega} J. M. Ortega and W. C. Rheinboldt, \textit{Iterative
Solution of Nonlinear Equations in Several Variables}, SIAM Ed., 2000.

\bibitem{Agarwal-book} R. P. Agarwal, M. Meehan, and D. O' Regan,
\emph{Fixed Point Theory and Application},\textquotedblright \textit{\emph{Cambridge
University Press}}, 2001. 

\bibitem{Bellman}S. N. Elaydi, \textit{An Introduction to Difference
Equations}, Springer, 3rd ed., 2005.

\bibitem{Larsson-Jorswieck}E. Larsson and E. Jorswieck, {}``Competition
and Collaboration on the MISO Interference Channel,\textquotedblright{}
in \emph{Proc. of Allerton Conference on Communication, Control, and
Computing, 2007}{} (invited paper).

\bibitem{Ye-Blum-SP}S. Ye and R. S. Blum, {}``Optimized Signaling
for MIMO Interference Systems With Feedback,\textquotedblright\textit{{}IEEE
Trans. on Signal Processing}, vol. 51, no. 11, pp. 2839-2848, November
2003.

\bibitem{Demirkol-Ingram_VTC01}M. F. Demirkol and M. A. Ingram, {}``Power-Controlled
Capacity for Interfering MIMO Links,\textquotedblright\textit{{}}
in \emph{Proc. of the IEEE Vehicular Technology Conference (VTC 2001),
2001.}

\bibitem{Liang-Dandekar-WC} C. Liang and K. R. Dandekar, {}``Power
Management in MIMO Ad Hoc Networks: A Game-Theoretic Approach,\textquotedblright\textit{{}}
\textit{IEEE Trans. on Wireless Communications}, vol. 6, no. 4, pp.
2866-2882, April 2007.

\bibitem{Arslan_etal}G. Arslan, M. Fatih Demirkol and Y. Song, {}``Equilibrium
efficiency improvement in MIMO interference systems: a decentralized
stream control approach,\textquotedblright\textit{{}} \emph{IEEE
Transaction on Wireless Communications}, vol. 6, no. 8, pp. 2984--2993,
August 2007.

\bibitem{Rosen}J. Rosen, {}``Existence and Uniqueness of Equilibrium
Points for Concave n-Person Games,\textquotedblright{} \emph{Econometrica},
vol. 33, no. 3, pp. 520\textendash{}534, July 1965.

\bibitem{Cover} T. M. Cover and J. A. Thomas, \textit{Elements of
Information Theory}, John Wiley and Sons, 1991.

\bibitem{Forney-91 }J. G. David Forney and M. V. Eyuboglu, {}``Combined
Equalization and Coding Using Precoding,\textquotedblright\textit{{}}
\emph{IEEE Comm. Magazine}, vol. 29, no. 12, pp. 25\textendash{}34,
Dec. 1991.

\bibitem{Khalil} H. K. Khalil, \textit{Nonlinear Systems}, \textit{Prentice
Hall, Third Ed., 2002.}

\bibitem{Liberzon-book} D. Liberzon, \textit{Switching in Systems
and Control}, Springer, 1993.

\bibitem{Campbell-Meyer-book}S. L. Campbell and C.D. Meyer, \emph{Generalized
Inverse of Linear Transformations,} Dover Publications, 1991. 

\bibitem{Bernstein} D. S. Bernstein, \emph{Matrix Mathematics: Theory,
Facts, and Formulas with Application to Linear Systems Theory, }Princeton
University Press, February 22, 2005.

\bibitem{CPStone92}R. W. Cottle, J.-S. Pang, and R. E. Stone, \textsl{The
Linear Complementarity Problem}, Academic Press (Cambridge 1992).

\bibitem{Facchinei} F. Facchinei and J.-S. Pang, \textit{Finite-Dimensional
Variational Inequalities and Complementarity Problems}, Springer,
2000.

\bibitem{Horn85} { R. A. Horn and C. R. Johnson, \textit{ Matrix
Analysis}, Cambridge Univ. Press, 1985.}

\bibitem{Boyd-book}S. Boyd and L. Vandenberghe, \emph{Convex Optimization},
Cambridge University Press, 2003.

\end{thebibliography}
\end{document}